\documentclass[a4paper,11pt]{article}
\pdfoutput=1

\usepackage{jheppub}
\usepackage[T1]{fontenc}
\usepackage[utf8]{inputenc}
\usepackage{amsmath,amssymb,bm}
\usepackage{makecell}
\usepackage{booktabs}
\usepackage{float}
\usepackage{array}
\usepackage{multirow}
\usepackage{enumitem}
\usepackage{url}
\allowdisplaybreaks

\title{Clock-noise subtraction in geometric time-delay interferometry for space-based gravitational-wave parameter estimation}

\author[a]{Rui Luo}
\author[b]{Pan-Pan Wang\textsuperscript{*}}
\author[a]{Zi-Jiang Yang}
\author[c,d]{Wei-Liang Qian}
\author[e]{Cheng-Gang Shao\textsuperscript{*}}

\affiliation[a]{National Gravitation Laboratory, MOE Key Laboratory of Fundamental Physical Quantities Measurement, and School of Physics, Huazhong University of Science and Technology, Wuhan 430074, People's Republic of China}
\affiliation[b]{College of Physics, Chongqing University, Chongqing 401331, China}
\affiliation[c]{Faculdade de Engenharia de Guaratinguet\'a, Universidade Estadual Paulista, 12516-410, Guaratinguet\'a, SP, Brazil}
\affiliation[d]{Escola de Engenharia de Lorena, Universidade de S\~ao Paulo, 12602-810, Lorena, SP, Brazil}
\affiliation[e]{School of Physics and Optoelectronic, Yangtze University, Jingzhou 434023, China}

\emailAdd{ppwang@cqu.edu.cn (corresponding author)}
\emailAdd{cgshao@hust.edu.cn (corresponding author)}

\abstract{Millihertz gravitational-wave observations with space-based interferometers require time-delay interferometry (TDI) observables whose residual instrumental noise is sufficiently controlled for both detection and parameter inference.
Although TDI suppresses laser phase noise in unequal and time-dependent arms, clock jitter from onboard ultra-stable oscillators can remain above the secondary-noise floor and bias the effective noise weighting used in data analysis.
We formulate a clock-noise subtraction scheme directly in the geometric-TDI framework.
The construction introduces generalized clock-noise observables for the four space-time link structures that arise when both delay and time-advance operators are allowed.
This makes the clock-noise residual algebraically parallel to the laser-noise residual and yields explicit subtraction terms for arbitrary two-path geometric TDI observables.
We illustrate the method with representative first- and second-generation geometric TDI combinations, and test it with time-domain simulations using LISA-like orbits and noise levels.
For a modified second-generation U-type observable, the subtraction suppresses the clock-noise residual below the signal region, restores the expected sensitivity to a monochromatic source, and improves the Fisher and Markov-chain Monte Carlo parameter constraints on the source amplitude, frequency and phase.
These results show that clock-noise calibration is a necessary component of precision data analysis for future space-based gravitational-wave detectors.}


\begin{document}
\maketitle

\section{Introduction}
\label{section1}

The millihertz gravitational-wave (GW) band is expected to contain signals from massive black-hole binaries, compact Galactic binaries, extreme-mass-ratio inspirals and stochastic backgrounds.
Space-based observatories such as LISA~\cite{gw-lisa1,gw-lisa2}, TianQin~\cite{gw-tianqin} and Taiji~\cite{gw-Taiji} are designed to access this band and to provide information on source formation, strong-field gravity and the astrophysical environments of compact objects.
Realizing this scientific potential requires more than detecting excess power in the data.
The observed time series must support phase-coherent inference of source parameters over long observation times, and this makes instrumental-noise calibration a central element of the data-analysis problem.

The dominant instrumental disturbance in a space-based interferometer is laser phase noise.
Because the constellation arms are unequal and vary with the spacecraft orbits, laser noise cannot be removed by a simple equal-arm Michelson subtraction.
Time-delay interferometry (TDI)~\cite{tdi-01,tdi-02,tdi-03} instead forms delayed combinations of one-way measurements that synthesize equal optical paths and suppress the laser contribution.
The development of TDI has proceeded along both theoretical~\cite{frame-01-2000,tdi-d55-2001,res-semi--01-2002,tdi-Algebraic-2002,tdi-d22,tdi-laser-01,tdi-d99,tdi-d88,tdi-laser-06,tdi-laser-LISACode,tdi-2010-Dhurandhar,tdi-otto-2015,tdi-filter-s4,algebra-tdi-Wu,algebra-tdi-Qian} and experimental~\cite{TDIexper-deVine-2010,TDIexper-Vinckier-2020} directions for more than two decades.
Among the available constructions, geometric TDI is particularly useful for organizing the problem: a valid observable is represented by the interference of two virtual optical paths in a space-time diagram~\cite{tdi-geometric-2005,tdi-geometric-2020,tdi-geometric-2021,tdi-geometric-2022}.
This representation is well suited to second-generation and modified second-generation combinations, including observables that contain time-advance operators.

After laser-noise suppression, clock jitter associated with onboard ultra-stable oscillators (USOs) can become a limiting residual.
For representative mission parameters, the uncompensated USO contribution can exceed the secondary-noise floor and obscure the GW response in the relevant frequency band~\cite{tdi-clock-2001,tdi-clock-2002,tdi-clock-2012,tdi-clock-2015,tdi-clock-2018,tdi-clock-2021,tdi-clock-pan}.
This issue is not only a matter of sensitivity-curve presentation.
Residual clock noise enters the effective noise power spectral density used in likelihood evaluations, changes signal-to-noise ratios and broadens parameter posteriors.
Clock-noise subtraction is therefore part of the same precision-data-analysis program as laser-noise cancellation.

Sideband-based clock-noise removal was first introduced for static interferometers~\cite{tdi-clock-1996}, extended to first-generation TDI combinations~\cite{tdi-clock-2001,tdi-clock-2002}, and later generalized to second-generation and geometric TDI observables~\cite{tdi-clock-2018,tdi-clock-2021}.
Related approaches include algebraic clock-noise calibration~\cite{tdi-clock-pan}, direct construction of USO-free second-generation observables~\cite{tdi-clock-2012}, and the use of optical frequency combs~\cite{tdi-clock-2015}.
The sideband scheme transfers clock fluctuations to the laser phase through electro-optic modulation.
Subtracting carrier and sideband measurements gives observables that encode differences between remote and local clock noises, which can then be delayed and recombined to remove the clock contribution from a TDI data stream.
Despite this operational picture, a compact prescription for arbitrary geometric TDI combinations remains useful, especially when time advances are present.
Such a prescription should make explicit which measured data streams are required, how the clock residual propagates along the two virtual paths and how the subtraction can be implemented algorithmically.

In this work we formulate clock-noise subtraction directly within the geometric-TDI framework.
We introduce generalized clock-noise observables associated with the four possible space-time link structures encountered when delay and advance operators are both allowed.
With these observables, the clock-noise residual can be written in a form that mirrors the laser-noise residual of a geometric TDI combination.
This analogy leads to explicit subtraction terms for arbitrary two-path geometric TDI observables.
We give representative examples, including the Monitor-E combination, the modified second-generation Michelson combination $[X]_1^{16}$ and the modified second-generation $[U]_3^{16}$ combination.
We then validate the construction with time-domain simulations using LISA-like orbits and noise levels, and compare the residual PSD and sensitivity before and after clock-noise subtraction.
Finally, to connect the calibration scheme with data analysis, we quantify the impact on parameter estimation for a monochromatic source observed through the $[U]_3^{16}$ observable.

The paper is organized as follows.
In Sec.~\ref{section2} we define the interferometric data streams, delay conventions and clock-noise observables.
In Sec.~\ref{section3} we derive the residual clock-noise equation and its cancellation in the geometric-TDI language.
In Sec.~\ref{section4} we discuss the algorithmic implementation and provide explicit TDI examples.
Numerical simulations are presented in Sec.~\ref{section5}.
In Sec.~\ref{section6} we discuss the impact of clock-noise subtraction on parameter estimation for TDI combinations, using $[U]_3^{16}$ as a representative example.
Concluding remarks are given in Sec.~\ref{section7}.

\section{Interferometric measurements}\label{section2}

\subsection{Notation and interferometric measurement data streams}\label{section2.1}

\begin{figure}[H]
	\centering
	\includegraphics[width=0.62\textwidth]{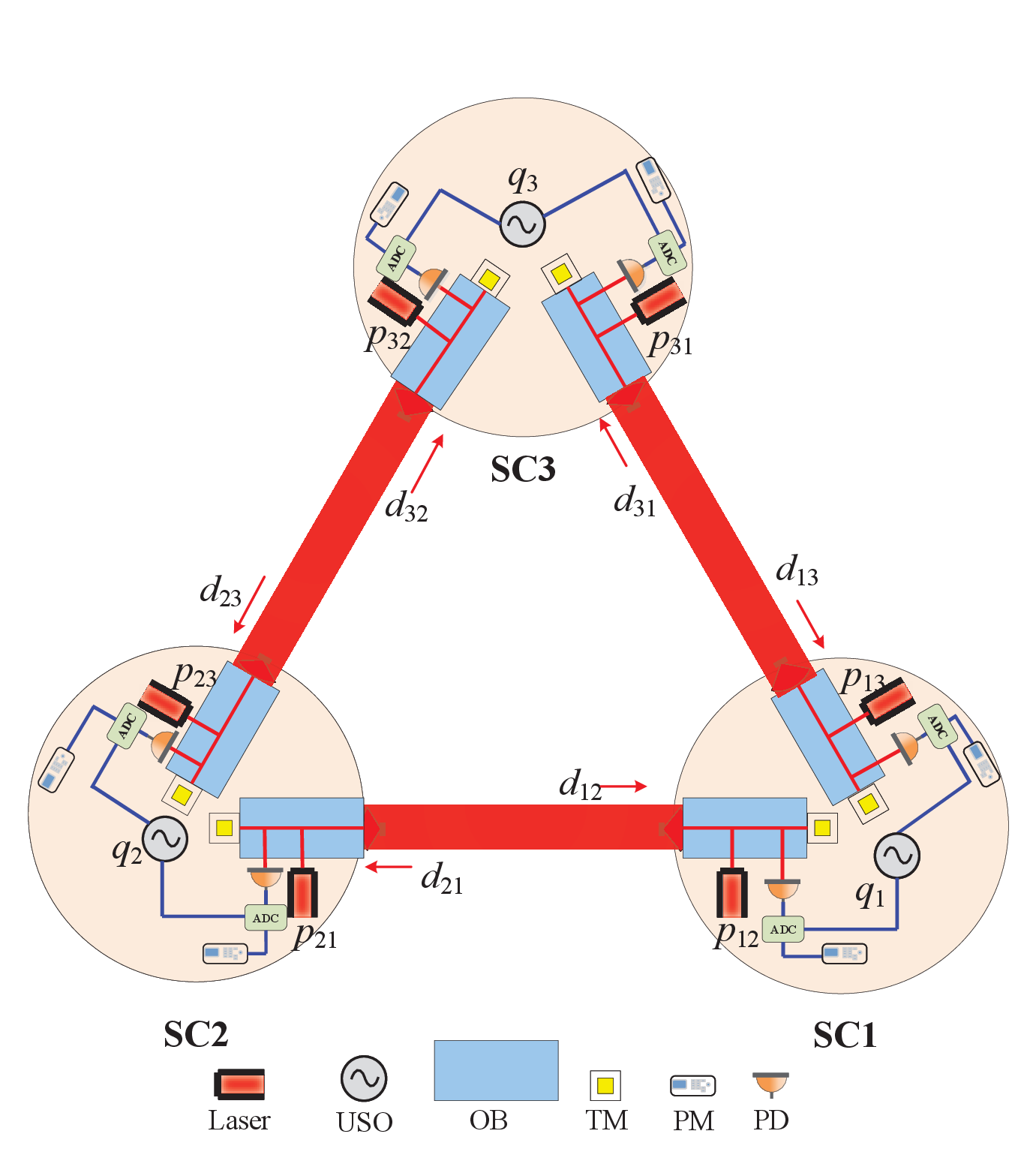}
	\caption{\label{fig1}
		The experimental layout of the constellation consists of three spacecraft and six optical benches.
		SC: spacecraft; USO: ultra-stable oscillators; OB: optical bench; TM:test mass; PM: phasemeters; PD: photodetector; ADC: analog-to-digital converter.}
\end{figure}

The space-based GW detector consists of three spacecraft labeled SC1, SC2, and SC3, respectively, in the clockwise direction depicted in Fig.~\ref{fig1}.
Each individual spacecraft $i$ possesses two optical benches (referred to as ``left'' and ``right''), which are indicated by the indices $ij$ and $ik$.
For optical benches' indices, the preceding index $i$ inherits from the local spacecraft, while the latter index $(j/k)$ indicates a second optical bench from a remote spacecraft, with whose light signal the interferometric measurements are performed.
The above double index notation will be prominently adopted in this paper\footnote{The above convention is utilized under two different contexts.
	First, it readily applies to any quantity associated with an interferometric measurement involving a pair of optical benches.
	Second, for a localized quantity, such as laser noise, the first index matches that of the spacecraft hosting the laser, while the second index indicates the remote spacecraft exchanging light with the local one.
	Subsequently, there are a total of six possibilities corresponding to six lasers.
	Throughout this paper, the index pair ``$ij$'' corresponds to a link $j\to i$ in the counterclockwise direction, namely, ``$i,(i+1)$'', while ``$ik$'' is used to imply that in the clockwise direction ``$i,(i-1)$''.}.
The relevant interferometric measurements are referred to as ``data streams'' in the literature.
Specifically, $s^{c}_{ij}(t)$ is the scientific carrier data stream, which represents the beat between laser beams of the local and remote optical benches $i$ and $j$, which potentially carries the GW signals.
$s^{sb}_{ij}(t)$ is the scientific sideband data stream, which contains the sideband interference pattern between the local and remote spacecraft.
$\varepsilon_{ij}(t)$ gives the test mass data stream, which measures the interferometric pattern between the two local laser beams while one of them bounces off from the local test mass.
$\tau_{ij}(t)$ is the reference data stream, which gives the beat between the two local laser beams without involving the test mass.
The above data streams possess the following forms:
\begin{subequations}
	\begin{align}
		s^{c}_{ij}(t) &= h_{ij}(t)+{\cal D}_{ij}p_{ji}(t)-p_{ij}(t)-a_{ij}q_{i}(t)\notag\\
		&+2\pi\nu_{ji}[{\cal D}_{ij}\vec{n}_{ij} \cdot\vec{\Delta}_{ji}(t)+\vec{n}_{ji}\cdot\vec{\Delta}_{ij}(t)]+N_{ij}^{{\rm{opt}}}(t),\label{sci}\\
		s^{sb}_{ij}(t) &= h_{ij}(t)\!+\!{\cal D}_{ij}p_{ji}(t)\!-\!p_{ij}(t)\!-\!c_{ij}q_{i}(t)\!+\!\nu^{m}_{ji}{\cal D}_{ij}q_{j}(t)\!-\!\nu^{m}_{ij}q_{i}(t)\notag\\
		&+2\pi\nu_{ji}[{\cal D}_{ij}\vec{n}_{ij} \cdot\vec{\Delta}_{ji}(t)+\vec{n}_{ji}\cdot\vec{\Delta}_{ij}(t)]+N_{ij}^{{\rm{opt}},sb}(t),\label{sb}\\
		\varepsilon_{ij}(t) &=p_{ik}(t)-p_{ij}(t)-b_{ij}q_{i}(t)+\mu_{ik}(t)\notag\\
		&-4\pi\nu_{ik}[\vec{n}_{ji} \cdot \vec{\delta}_{ij}(t)-\vec{n}_{ji}\cdot\vec{\Delta}_{ij}(t)],\label{test}\\
		\tau_{ij}(t) &=p_{ik}(t)-p_{ij}(t)-b_{ij}q_{i}(t)+\mu_{ik}(t),\label{ref}
	\end{align}
\end{subequations}
where $i,j,k\in \{1,2,3\}$\footnote{It is noted that the indices must take different values when they appear in pairs as variable subscripts. 
	For all the expressions considered in this paper, all possible cyclic permutations among the three distinct values do not affect the validity of the expression.
	However, the expression may (e.g., Eq.~\eqref{ref}) or may not (e.g., Eqs.~\eqref{etaijik}, ~\eqref{xiclock} and~\eqref{delta1}) remain valid when one exchanges two distinct indices while holding the third one unchanged.}, $h_{ij}$ represents the contributions from the incident GWs, $p_{ij}$ indicates the laser phase noise\footnote{According to the double index convention, $p_{12}$ represents the laser phase noise of the laser on spacecraft $1$ that exchanges light with its counterpart on spacecraft $2$, while $p_{21}$ is the laser phase noise of the latter laser on spacecraft $2$.
	In comparison, $s^c_{12}$ corresponds to the link $2\to 1$, indicating that the trajectory of the light beam is in the counterclockwise direction, which points to the right in the corresponding space-time diagram.
	Similarly, $s^c_{21}$ represents the link $1\to 2$ whose trajectory is in the clockwise direction, which points to the left in the corresponding space-time diagram.}, $q_{i}$ gives the clock-jitter noise due to the USO fluctuations on spacecraft $i$, $\vec{n}\cdot\vec{\delta}$ and $\vec{n}\cdot\vec{\Delta}$ are the contributions from the test mass and optical bench noises respectively, the unit vectors $\vec{n}_{ij}$ are along the directions of the propagation of the laser beams, in a counterclockwise fashion,
$\nu_{ij}$ is the carrier frequencies, $\nu^{m}_{ij}$ gives sideband frequency offset, $N_{ij}^{{\rm{opt}}}$ are optical path noise, and $\mu_{ik}$ are the optical fibers noise.
The time delay operator ${\cal D}_{ij}$ is defined as
\begin{align}\label{delaydefine}
	{\cal D}_{ij} x( t)=x\left ( t-d _{ij}( t) \right ),
\end{align}
where $d_{ij}(t)$ represents the light travel time for spacecraft $i$ receiving the light signal emitted from spacecraft $j$ (where the speed of light $c$ is assumed to be unity).
As a function of time governed by the satellites' orbits, $d_{ij}(t)$ can be expanded as
\begin{align}\label{ddefine}
	d_{i j}(t)={d}_{i j}+\dot{d}_{i j} t+\frac12\ddot{d}_{i j} t^{2} +\ldots,
\end{align}
where ${d}_{i j}$, $\dot{d}_{i j}$ and $\ddot{d}_{i j}$ are the armlength, their first-order and second-order derivatives at $t=0$, respectively.
For the cascaded time-delay operators, one introduces a simple notation 
\begin{align}\label{delaynonation}
	{\cal D}_{i_1 i_2\cdots i_n} \equiv {\cal D}_{i_1 i_2}{\cal D}_{i_2 i_3}{\cal D}_{i_3 i_4}\cdots {\cal D}_{i_{n - 2} i_{n - 1}}{\cal D}_{i_{n - 1}{i_n}}.
\end{align}
We denote the time-advance operator as ${\cal D}_{-ji}$\footnote{The negative sign before the pair of indeces $ji$ should be interpreted as ``time-advance''. 
	As it will become apparent later, we adopt this convention as it conveniently provides explicit information on successive links joining together to form a continuous trajectory in the space-time diagram. Besides, neither the superscripted form ${\cal D}_{ij}^{-1}$ nor an overline $\overline{{\cal D}_{ij}}$ or ${\cal D}_{\overline{ij}}$ will do a better job when applying to the forthcoming expressions such as Eqs.~\eqref{intermeres},~\eqref{intereta}, and~\eqref{clockkth}.}, which is defined as
\begin{align}\label{advancedefine}
	{\cal D}_{ - ji}{\cal D}_{ij}x(t) = {\cal D}_{ij}{\cal D}_{ - ji}x(t) = x(t) .
\end{align}
The specific form of the time-advance operator can be taken as
\begin{align}
	{\cal D}_{-ji} x( t)&=x\left ( t+d _{ij}(t+d _{ij})\right ),
\end{align}
where one has neglected the quadratic and higher-order terms.
The three coefficients $a$, $b$, and $c$ are essentially beat-note frequencies between the laser beams, given as
\begin{align}\label{abcij}
	a_{ij}&=(1- \dot{d}_{i j})\nu_{j i}-\nu_{i j},\notag\\
	b_{ij}&=\nu_{i k}-\nu_{i j},\notag\\
	c_{ij}&=(1- \dot{d}_{i j})(\nu_{j i}+\nu^{m}_{j i})-(\nu_{i j}+\nu^{m}_{i j}) .
\end{align}
In literature~\cite{tdi-clock-2001, tdi-clock-2018}, $q_{i}$ can be alternatively defined as the phase noise due to the clock fluctuations onboard spacecraft $i$, thus the coefficients ($a$, $b$, and $c$) become dimensionless, defined as the above beat-note frequencies divided by the constant USO-referenced pilot-tone frequency.
It is noted that the difference in the definitions does not affect the clock-noise reduction algorithm.

\subsection{Intermediary variables}\label{section2.2}

It is convenient to introduce some intermediate variables to suppress the laser phase noise, and optical bench noise contained in the interferometric data streams.

Firstly, the intermediary variables $\xi_{ij}(t)$ can be constructed to eliminate the optical benches noise~\cite{tdi-03}:
\begin{align}\label{xiij}
	\xi_{ij}(t)=s^{c}_{ij}(t)-\frac{\nu_{ji}}{\nu_{ik}}\frac{\varepsilon_{ij}(t)-\tau_{ij}(t)}{2}-\frac{\nu_{ji}}{\nu_{jk}}{\cal D}_{ij}\frac{\varepsilon_{ji}(t)-\tau_{ji}(t)}{2}.
\end{align}
Subsequently, the intermediate variables $\eta_{ij}$ or $\eta_{ik}$ are constructed to furnish the TDI solutions for the primed laser phase noise:
\begin{subequations}
	\begin{align}
		\eta_{ij}(t)=&\xi_{ij}(t)-{\cal D}_{ij}\frac{\tau_{jk}(t)-\tau_{ji}(t)}{2},\label{etaaij}\\
		\eta_{ik}(t)=&\xi_{ik}(t)+\frac{\tau_{ij}(t)-\tau_{ik}(t)}{2} ,\label{etaaik}
	\end{align}
\end{subequations}
where for the indices, the tuple $(i, j, k)$ is defined to be in cyclic order.

By substituting Eqs.~\eqref{sci},~\eqref{test},~\eqref{ref}, and~\eqref{xiij} into Eqs.~\eqref{etaaij},~\eqref{etaaik}, the intermediary variables $\eta_{ij}$ or $\eta_{ik}$ can be rewritten as
\begin{subequations}
	\begin{align}
		\eta_{ij}(t)&=h_{ij}(t)+{\cal D}_{ij}p_{j}(t)-p_{i}(t)+b_{jk}{\cal D}_{ij}q_{j}(t)-a_{ij}q_{i}(t)\notag\\
		&+2\pi\nu_{ji}[{\cal D}_{ij}\vec{n}_{ij} \cdot \vec{\delta}_{ji}(t)+\vec{n}_{ji}\cdot\vec{\delta}_{ij}(t)]+N_{ij}^{{\rm{opt}}}(t),\label{etaij}\\
		\eta_{ik}(t)&=h_{ik}(t)+{\cal D}_{ik}p_{k}(t)-p_{i}(t)-(b_{ij}+a_{ik})q_{i}(t)\notag\\
		&+2\pi\nu_{ki}[{\cal D}_{ik} \vec{n}_{ik} \cdot \vec{\delta}_{ki}(t)+\vec{n}_{ki}\cdot\vec{\delta}_{ik}(t)]+N_{ik}^{{\rm{opt}}}(t).\label{etaik}
	\end{align}\label{etaijik}
\end{subequations}
This combined data stream contains only the three laser phase noise $p_{ij}(t)$ from one of the two optical benches. 
Therefore, we denote it by a single index $p_{i}(t)\equiv p_{ij}(t)$. 
For the sake of simplicity, we will also drop the explicit time dependency from now on.

To deal with the clock noise, one subtracts the sideband data streams $s^{sb}_{ij}$ from the carrier data streams $s^{c}_{ij}$ to define additional variables.
The resulting six intermediate variables $r_{ij}$ and their generalizations elaborated below in Sec.~\ref{section3.2} will be referred to as {\it clock-noise observable}. 
In the literature~\cite{tdi-03, tdi-clock-2001, tdi-clock-2018}, they were originally constructed, via Eqs.~\eqref{sci} and~\eqref{sb}, to have the following forms
\begin{align}\label{rrij}
	r_{ij}\equiv\frac{s^{sb}_{ij}-s^{c}_{ij}}{\nu^{m}_{j i}}={\cal D}_{i j}q_{j}-q_{i} .
\end{align}
We note that these variables only concern the clock noise, where the Doppler effect is ignored.

\section{General scheme for clock noise cancellation}\label{section3}

\subsection{Residual laser phase noise of an arbitrary geometric TDI solution}\label{section3.1}

\begin{figure}[H]
	\centering
	\includegraphics[width=0.2\textwidth]{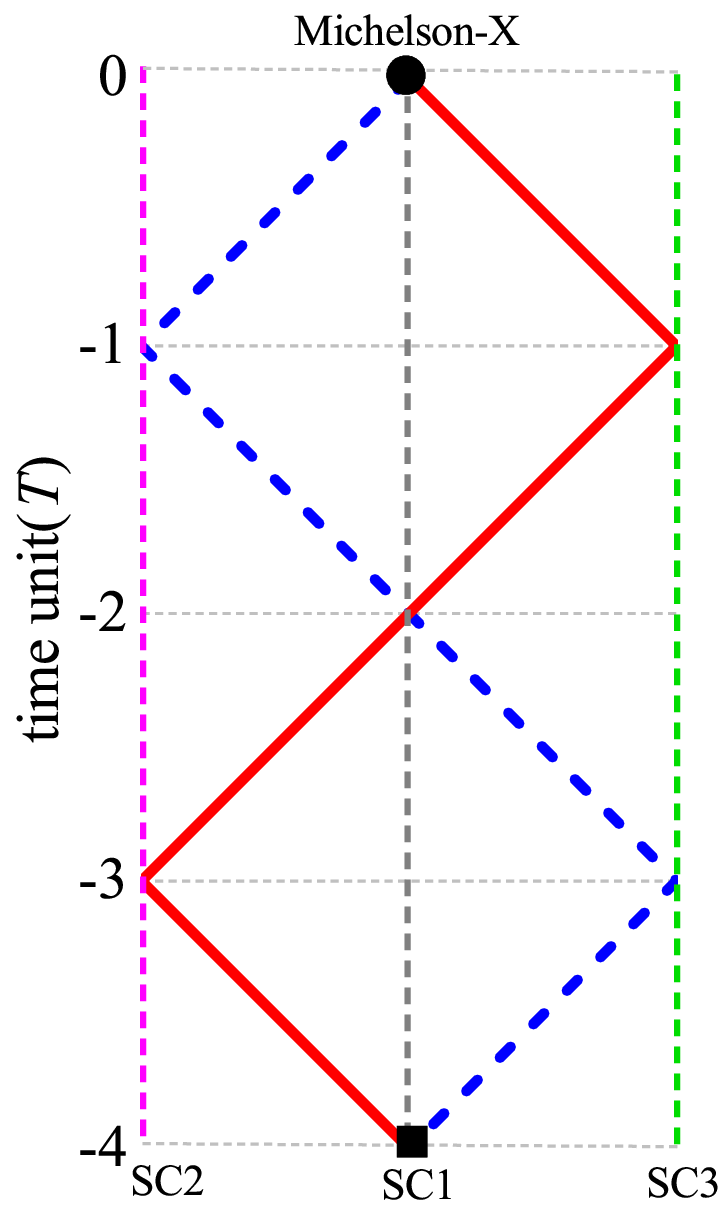}\hspace{20pt}
	\includegraphics[width=0.14\textwidth]{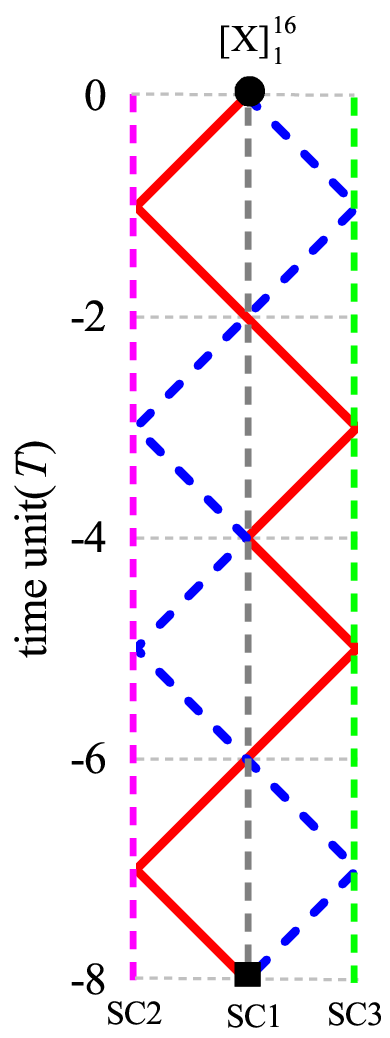}
	\caption{\label{fig2}
		The space-time diagram of the eight-link Michelson combination $X_1$ (left) and the sixteen-link Michelson combination $[X]_1^{16}$ (right).
		The red-solid and blue-dashed line segments correspond to the two synthesized routes, which might contain both forward and backward propagation in time.
		The black squares and black dots indicate the initial and terminal grids of the two routes, and the reference time $t=0$ is attributed to the terminal SC 1.}
\end{figure}

Let us concentrate on the TDI solutions that can be derived using the geometric TDI algorithm.
A typical geometric TDI solution can be presented using the space-time diagram, which involves two virtual equal-arm optical paths~\footnote{In this paper, we consider a geometric TDI solution consisting of two virtual optical paths, both with $n$ links.
	Although, in general, the two optical paths do not necessarily possess the same number of links, the above assumption does not undermine the generality of our findings, as it still suffices to exhaust all possible geometric TDI combinations.}, as shown in Fig.~\ref{fig2}.
Intuitively, when focusing on the laser phase noise, the residual for a given geometric TDI solution can be expressed as the difference:
\begin{align}\label{georesidual}
	\mathrm{TDI}_\mathrm{geo}^{p}&=({\cal D}_{\pm i_0 i_1} {\cal D}_{\pm i_1 i_2} \ldots{\cal D}_{\pm i_{n-2} i_{n-1}}{\cal D}_{\pm i_{n-1} i_n}\nonumber\\
	&-{\cal D}_{\pm i_{0} i'_1}{\cal D}_{\pm i'_1 i'_2} \ldots{\cal D}_{\pm i'_{n-2} i'_{n-1}} {\cal D}_{\pm i'_{n-1} i_n}) p_{i_n},
\end{align}
where the subscripts of the time-displacement operators are the indices of spacecraft, the plus (minus) sign before a pair of indices indicates that it is a time-delay (time-advance) operator, and the indices with prime $(')$ indicate that the contributions are from the links on the second virtual optical path.
Eq.~\eqref{georesidual} can be further simplified as
\begin{align}\label{geoexpression}
	\mathrm{TDI}_\mathrm{geo}^{p}= \left( {\prod\limits_{k = 1}^n {{\cal D}_{ \pm {i_{k - 1}}{i_k}}}  - \prod\limits_{k = 1}^n {{\cal D}_{ \pm {i'_{k - 1}}{i'_k}}} } \right){p_{{i_n}}},
\end{align}
where $i_0=i'_0$ and $i_n=i'_n$.

For example, the space-time diagram of the modified first-generation Michelson combination $X_1$ is shown in the left plot of Fig.~\ref{fig2}.
Emitted from spacecraft 1, the laser beams propagate along two different virtual paths $1\to 2\to 1\to 3\to 1$ and $1\to 3\to 1\to 2\to 1$, finally interfering at spacecraft 1.
The corresponding residual reads
\begin{align}\label{Xresdiual}
	\mathrm{TDI}_{X_1}^p = ({\cal D}_{13}{\cal D}_{31}{\cal D}_{12}{\cal D}_{21} - {\cal D}_{12}{\cal D}_{21}{\cal D}_{13}{\cal D}_{31})p_1 .
\end{align}

As a second example, in the right plot of Fig.~\ref{fig2}, we show the space-time diagram of the modified second-generation Michelson combination $[X]_1^{16}$.
Similarly, the residual is found to be
\begin{align}\label{X2resdiual}
	\mathrm{TDI}_{[X]_1^{16}}^p &= ({\cal D}_{12}{\cal D}_{21}{\cal D}_{13}{\cal D}_{31}{\cal D}_{13}{\cal D}_{31}{\cal D}_{12}{\cal D}_{21}\nonumber\\
	&- {\cal D}_{13}{\cal D}_{31}{\cal D}_{12}{\cal D}_{21}{\cal D}_{12}{\cal D}_{21}{\cal D}_{13}{\cal D}_{31})p_1.
\end{align}

In what follows, we derive a general form of the residual noise from an arbitrary geometric TDI solution.
To proceed, one subtracts $p_{i_0}$ from the first term of Eq.~\eqref{georesidual} to find
\begin{align}\label{intermeres}
	&{\cal D}_{ \pm i_0 i_1}{\cal D}_{ \pm i_1 i_2}\cdots{\cal D}_{ \pm i_{n - 2}i_{n-1}}{\cal D}_{ \pm i_{n - 1}i_n}p_{i_n} - p_{i_0}\nonumber\\
	=& \left( \prod\limits_{k = 1}^n {\cal D}_{ \pm i_{k - 1}i_k} \right)p_{i_n} - p_{i_0}\nonumber\\
	=& {\cal D}_{ \pm {i_0}i_1\cdots i_{n - 2}i_{n - 1}}({\cal D}_{ \pm i_{n - 1}{i_n}}p_{i_n} \!-\! p_{i_{n - 1}}) \!+\! {\cal D}_{ \pm {i_0}i_1\cdots i_{n - 3}i_{n - 2}}({\cal D}_{ \pm i_{n - 2}i_{n - 1}}p_{i_{n - 1}} \!-\! p_{i_{n - 2}})\nonumber\\
	+& \cdots+{\cal D}_{ \pm {i_0}{i_1}}({\cal D}_{ \pm {i_1}{i_2}}{p_{i_2}} - {p_{i_1}}) + ({\cal D}_{ \pm {i_0}{i_1}}{p_{i_1}} - {p_{i_0}}),
\end{align}
where for the last equality, each parenthesis contains two terms, the second term of the proceeding parenthesis precisely cancels out the first term from the succeeding parenthesis, and as a result, only the very first and very last terms remain.

The above residual can now be expressed in terms of the intermediate variables $\eta_{ij}$ defined in Eq.~\eqref{etaijik}.
For the TDI solution, we only need to retain the terms involving laser phase noise, namely, 
\begin{align}\label{etaijp}
	\eta^p_{ij}&={\cal D}_{ij}p_{j}-p_{i} .
\end{align} 
Moreover, a valid geometric TDI solution Eq.~\eqref{georesidual} might contain time-advance operators.
Specifically, as elaborated further below, a relevant solution is constructed by piecing together terms of the form
\begin{align}\label{etaijpGEN}
	{\cal D}_{\pm ij}p_{j}-p_{i} .
\end{align} 
To this end, we also introduce the variable $\eta^p_{ - ij}$ as follows
\begin{align}\label{etaijpadvan}
	\eta^p_{ - ij} = {\cal D}_{ - ij}p_j - p_i =  - {\cal D}_{ - ij}({\cal D}_{ji}p_i - p_j) =  - {\cal D}_{ - ij}\eta^p_{ji}.
\end{align} 

A few comments are in order before proceeding further.
The above derivation of Eq.~\eqref{etaijpadvan} was carried out from the viewpoint of the geometric TDI solution.
However, another level of subtlety involving the data streams perspective will become relevant when one generalizes Eq.~\eqref{etaijpadvan} to deal with clock-noise residual below in Sec.~\ref{section3.2}.
To be specific, Eq.~\eqref{etaijpGEN} must be constructed using measured data streams defined by Eqs.~\eqref{sci},~\eqref{sb},~\eqref{test}, and~\eqref{ref}, which involve the time-delay operations exclusively.
The definition of the generalized intermediate variable $\eta_{-ij}$, in turn, must take this into account.
As a result, one defines
\begin{align}\label{etaijpadvanGEN}
	\eta_{ - ij} =  - {\cal D}_{ - ij}\eta_{ji} ,
\end{align} 
whose r.h.s. ensures it can be constructed by the data streams according to Eq.~\eqref{etaijik}, while the l.h.s. guarantees that the terms pertaining to the laser noise serve as the building block for geometric TDI.

By making use of Eqs.~\eqref{etaijp} and~\eqref{etaijpadvan}, the r.h.s. of Eq.~\eqref{intermeres} can be written as
\begin{align}\label{intereta_laser_terms}
	&{\cal D}_{ \pm {i_0}i_1\cdots i_{n - 2}i_{n - 1}}\eta^p_{ \pm i_{n - 1}{i_n}} + {\cal D}_{ \pm {i_0}i_1\cdots i_{n - 3}i_{n - 2}}\eta^p_{ \pm i_{n - 2}i_{n - 1}}\nonumber\\
	&+\cdots  + {\cal D}_{ \pm {i_0}{i_1}}\eta^p_{ \pm {i_1}{i_2}} + \eta^p_{ \pm {i_0}{i_1}}.
\end{align}
Subsequently, Eq.~\eqref{georesidual} gives
\begin{align}\label{interetaLASER}
	\mathrm{TDI}_\mathrm{geo}^p &= (\eta^p_{ \pm {i_0}{i_1}} + {\cal D}_{ \pm {i_0}{i_1}}\eta^p_{ \pm {i_1}{i_2}}\nonumber\\
	&+ \cdots  + {\cal D}_{ \pm {i_0}{i_1}}{\cal D}_{ \pm {i_1}{i_2}}\cdots {\cal D}_{ \pm {i_{n - 3}}{i_{n - 2}}}{\cal D}_{ \pm {i_{n - 2}}{i_{n - 1}}}\eta^p_{ \pm {i_{n - 1}}{i_n}})\nonumber\\
	& - (\eta^p_{ \pm {i_0}{i'_1}} + {\cal D}_{ \pm {i_0}{i'_1}}\eta^p_{ \pm {i'_1}{i'_2}}\nonumber\\
	&+ \cdots + {\cal D}_{ \pm {i_0}{i'_1}}{\cal D}_{ \pm {i'_1}{i'_2}}\cdots {\cal D}_{ \pm {i'_{n - 3}}{i'_{n - 2}}}{\cal D}_{ \pm {i'_{n - 2}}{i'_{n - 1}}}\eta^p_{ \pm {i'_{n - 1}}{i_n}})\nonumber\\
	&= \sum\limits_{k = 1}^n {\left\{ {\left( {\prod\limits_{s = 1}^{k - 1} {\cal D}_{ \pm {i_{s - 1}}{i_s}} } \right)\eta^p_{ \pm {i_{k - 1}}{i_k}} - \left( {\prod\limits_{s = 1}^{k - 1} {\cal D}_{ \pm {i'_{s - 1}}{i'_s}} } \right)\eta^p_{ \pm {i'_{k - 1}}{i'_k}}} \right\}}.
\end{align} 
This is the expression for residual laser noise of a given geometric TDI combination.

It is noted that the above derivation was performed by primarily focusing on the laser phase noise. 
However, by definition, the intermediate variables $\eta_{\pm ij}$ also carry the remaining noises.
In other words, these results can be readily generalized to derive the residual clock noise of an arbitrary geometric TDI solution by simply replacing $\eta^p$ defined by Eq.~\eqref{etaijp} by the general form $\eta$ defined by Eq.~\eqref{etaijik}.
To be specific, besides Eq.~\eqref{interetaLASER}, the residual noise for an arbitrary geometric TDI solution reads
\begin{align}\label{intereta}
	\mathrm{TDI}_\mathrm{geo}^\mathrm{laser} 
	&= \sum\limits_{k = 1}^n {\left\{ {\left( {\prod\limits_{s = 1}^{k - 1} {\cal D}_{ \pm {i_{s - 1}}{i_s}} } \right)\eta_{ \pm {i_{k - 1}}{i_k}} - \left( {\prod\limits_{s = 1}^{k - 1} {\cal D}_{ \pm {i'_{s - 1}}{i'_s}} } \right)\eta_{ \pm {i'_{k - 1}}{i'_k}}} \right\}}.
\end{align} 
This result will be further explored in the following subsection.

Before closing this subsection, we note that a valid geometric TDI solution requires the two virtual optical paths to be identical.
The lowest order term of Eq.~\eqref{ddefine} implies the relation
\begin{align}\label{dequalarm}
	\sum\limits_{k = 1}^n d_{ \pm i_{k-1} i_k}-\sum\limits_{k = 1}^n d_{ \pm i_{k-1}' i_k'}=0,
\end{align} 
where $d_{-ij}=-d_{ji}$.
Specifically, Eq.~\eqref{dequalarm} must be satisfied by the modified first-generation, second-generation, and modified second-generation geometric TDI solutions.

\subsection{Residual clock noise and its cancellation equation}\label{section3.2}

\begin{figure}[H]
	\centering
	\includegraphics[width=0.60\textwidth]{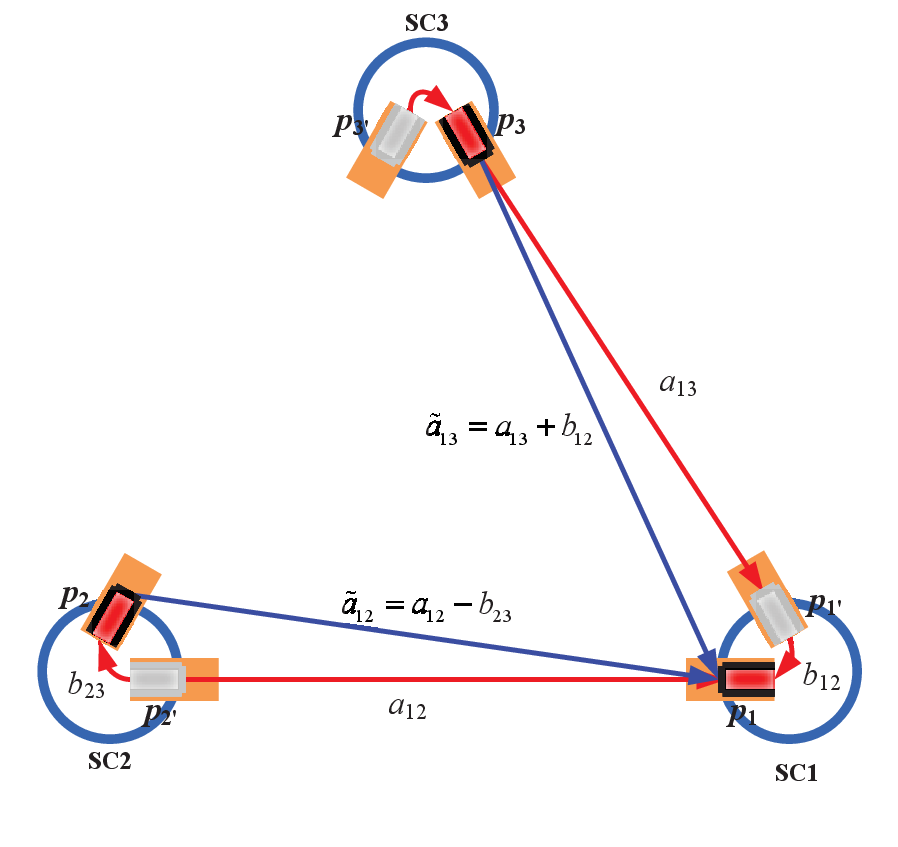}
	\caption{\label{fig3}
		Equivalent beat-note frequencies are illustrated as vector sums.
		The arrowed line segments in red are the original beat-note frequencies defined in Eqs.~\eqref{abcij}, and the effective ones given in Eqs.~\eqref{xiclockACof} are indicated by arrowed line segments in blue.}
\end{figure}

By exploiting the specific form of Eq.~\eqref{intereta}, in this subsection, we turn to discuss the residual clock noise and focus on the specific equation designed to eliminate that noise.
To proceed, we notice that one can recuperate the form of the clock noise in the noise residual Eq.~\eqref{intereta} by picking them out explicitly from the variable $\eta$ defined by Eqs.~\eqref{etaijik}.
To be specific, by picking out the terms exclusively pertaining to the clock noise, we have
\begin{align}
	\eta _{ij}^q&=b_{jk}{\cal D}_{ij}q_j-a_{ij}q_i =-\tilde a_{ij}q_i+b_{jk}r_{ij} = \tilde \eta _{ij}^q+\delta \eta _{ij}^q ,\label{etaClockij}\\
	\eta _{ik}^q&= - ({b_{ij}} + {a_{ik}}){q_i}=  -\tilde a_{ik}q_i = \tilde \eta _{ik}^q,\label{etaClockik}
\end{align} 
where one introduces
\begin{align}\label{xiclock}
	\tilde \eta _{ij}^q &= -\tilde a_{ij}q_i,\quad \tilde \eta _{ik}^q=  -\tilde a_{ik}q_i,  \\
	\delta \eta_{ij}^q &= b_{jk}r_{ij}, \quad \delta \eta_{ik}^q = 0, \label{delta1}
\end{align}
and
\begin{align}\label{xiclockACof}
	\tilde a_{ij} = a_{ij}-b_{jk} ,\\\notag
	\tilde a_{ik}={b_{ij}} + {a_{ik}},
\end{align}
where $\tilde a_{ij}, \tilde a_{ik}$ represent equivalent beat-note frequencies after eliminating three of the six laser noise in Eqs.~\eqref{etaijik}.
Specifically, as shown in Fig.~\ref{fig3}, $a_{12}$ represents the beat-note frequency of the lasers $p_1$ and $p_{2'}(p_{21})$, $b_{23}$ is the beat-note frequency between the lasers $p_2$ and $p_{2'}(p_{21})$, and $\tilde a_{12}$ gives the effective beat-note frequency between the lasers $p_1$ and $p_2$.
Owing to its explicit form, $\delta \eta_{ij}^q$ is an observable that can be obtained from measured data streams.

We anticipate that the trailing term $b_{jk}r_{ij}$ will not pose any difficulty because its contribution to the noise residual is an explicit function of the observable $r_{ij}$, as will become apparent below.
Therefore, we will primarily concentrate on the contributions from $\tilde \eta _{ij}^q$ to the resultant clock-noise residual.

Moreover, to accommodate the ``time advance'' operations, we separate the terms relevant to the clock noise from Eq.~\eqref{etaijpadvanGEN} and find
\begin{align}\label{eta-ij-FULL}
	\eta _{ - ij}^q 
	= -{\cal D}_{ - ij}\eta^q_{ji}
	= -{\cal D}_{ - ij}\left(\tilde \eta_{ji}^q+\delta \eta_{ji}^q\right)  
	= \tilde \eta _{ - ij}^q+\delta \eta_{-ij}^q.
\end{align}
\begin{align}\label{xiclockinv}
	\tilde a _{ - ij}=  - \tilde a _{ji} ,
\end{align}
and introduces
\begin{align}\label{eta-ij}
	\tilde \eta _{ - ij}^q = {\cal D}_{ - ij}\tilde a _{ji}{q_j} = -\tilde a _{ - ij}{\cal D}_{ - ij}{q_j} \equiv -\tilde a_{ - ij}q_{ - i} ,
\end{align}
and
\begin{align} \label{delta2}
	\delta \eta_{-ij}^q =-{\cal D}_{ - ij} \delta \eta_{ji}^q,
\end{align}
where for Eq.~\eqref{eta-ij}, one utilizes Eq.~\eqref{xiclock} for the second equality, one makes use of Eq.~\eqref{xiclockinv} and the fact that $-\tilde a_{ - ij}$ is a number for the third equality, and the variable $q_{-i}$ with a ``negative'' subscript is defined in the last step to denote a clock noise propagated backwardly in time\footnote{It is important to note that $q_{-i}$ is defined in the context of a given node in the space-time diagram of a well-known geometric TDI solution, and therefore, there is no ambiguity about its preceding node $j$.}.
Like $\delta\eta_{ij}^q$, the quantity $\delta \eta_{-ij}^q$ is also an observable that can be obtained from measured data streams.
It is not difficult to show that the Eq.~\eqref{xiclock} is valid for both the pairs $ij$ and $ik$, and therefore, combining Eqs.~\eqref{xiclock} and~\eqref{eta-ij} give
\begin{align}\label{clockuniform}
	\tilde \eta _{ \pm ij}^q = -\tilde a _{ \pm ij}q_{ \pm i}.
\end{align}

Before proceeding to evaluate clock residual noise in more detail, besides the clock-noise observables $r_{ij}$ defined in Eq.~\eqref{rrij}
\begin{align}
	r_{ij} \equiv r_{i,j} = {\cal D}_{ij}{q_j} - {q_i}, \nonumber
\end{align}
we generalize the definition to consider negative indices by introducing another three intermediate variables of the form
\begin{align}
	r_{\pm i, \pm j}  .\nonumber
\end{align}
Specifically, one defines
\begin{align}\label{r-ijdata}
	r_{ - i,j} = {\cal D}_{ - ij}{q_j} - q_{ - i}= {\cal D}_{ - ij}{q_j} - {\cal D}_{ - ij}{q_j} = 0,
\end{align}
\begin{align}\label{ri-jdata}
	r_{i, - j}=& {\cal D}_{ij}q_{ - j} - q_i= {\cal D}_{ij}{\cal D}_{ - jk}{q_k} - {q_i}\notag\\
	=&  - {\cal D}_{ij}{\cal D}_{ - jk}({\cal D}_{kj}{q_j} - {q_k}) + {\cal D}_{ij}{q_j} - {q_i}\notag\\
	=&  - {\cal D}_{ij}{\cal D}_{ - jk}r_{kj} + r_{ij},
\end{align}
and
\begin{align}\label{r-i-jdata}
	r_{ - i, - j} =& {\cal D}_{ - ij}q_{ - j} - q_{ - i} = {\cal D}_{ - ij}{\cal D}_{ - jk}{q_k} - {\cal D}_{ - ij}{q_j}\notag\\
	=&  - {\cal D}_{ - ij}{\cal D}_{ - jk}({\cal D}_{kj}{q_j} - {q_k})\notag\\
	=&  - {\cal D}_{ - ij}{\cal D}_{ - jk}r_{kj}.
\end{align}

\begin{figure}[t]
	\centering
	\includegraphics[width=0.92\textwidth]{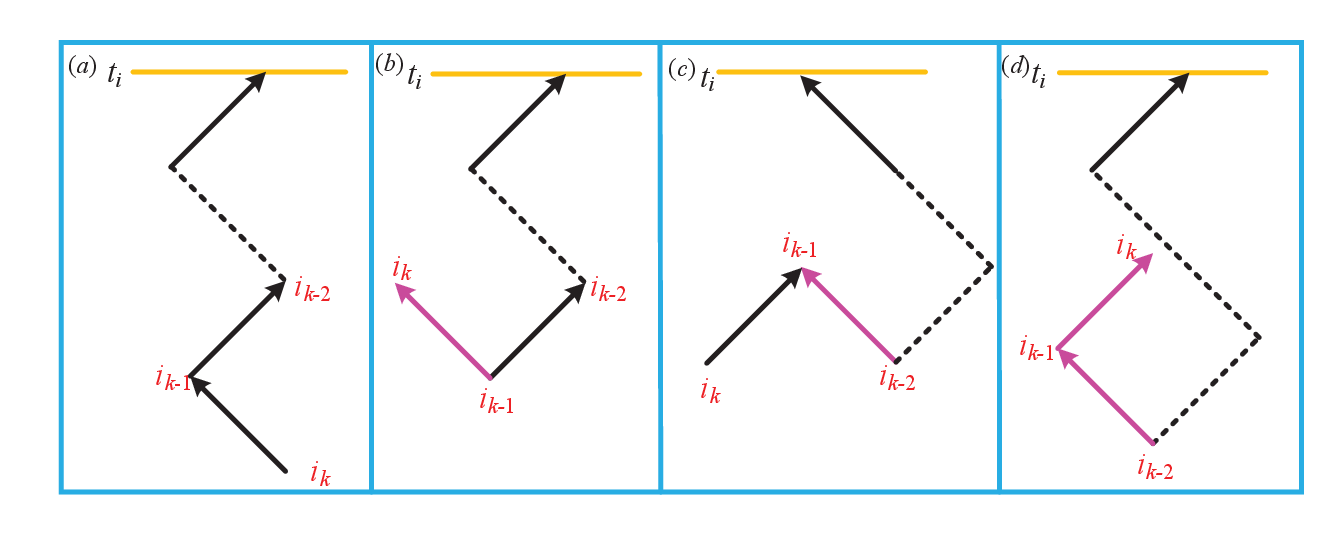}
	\caption{\label{fig4}
		Noise propagation in the space-time diagram for four exhaustive scenarios (a-d) for a given geometric TDI solution.
		The laser propagations in question primarily concern the link $i_{k-1}i_{k}$, but the preceding one $i_{k-2}i_{k-1}$ is also involved.
		The arrowed line segments in black indicate the propagations along the time direction, and the ones in magenta show those in the opposite direction of time.}
\end{figure}

It is observed that, in the definitions, certain subscripts have been assigned negative values,  corresponding to their associated virtual optical trajectories. 
As we elaborate further, the three quantities defined in Eqs.~\eqref{r-ijdata},~\eqref{ri-jdata}, and~\eqref{r-i-jdata} bear close relevance to the subfigures~(c),~(b), and~(d) depicted in Fig.~\ref{fig4}. 
Specifically, for subsequent discussions, we consider $k\equiv i_k$, $j\equiv i_{k-1}$, and $i\equiv i_{k-2}$.

We first consider the variable $r_{ - i,j}$ defined in Eq.~\eqref{r-ijdata}.
It corresponds to the noise $q_{j}$ propagating in the opposite direction of time to the satellite $i$, and then subtracted by the clock noise $q_{-i}$.
Such a term originated from the data $\tilde\eta^q_{jk}$, the part of $\eta_{jk}$ related to the clock noise, propagates from $j$ to $i$ in the opposite direction of time, by applying ${\cal D}_{ - ij}$, where by definition, the data associated with the second node $i$ is $\tilde\eta^q_{-ij}$.
Owing to Eq.~\eqref{clockuniform}, $\tilde\eta^q_{jk}$ and $\tilde\eta^q_{-ij}$ are proportional to, respectively, $q_j$ and $q_{-i}$.
It corresponds to the subfigure (c) shown in Fig.~\ref{fig4}.
Even though the process itself is not trivial, as it gives a non-vanishing contribution to the laser noise, it leads to no contribution in terms of the residual clock noise and might be cast away in the latter context.

For the second variable $r_{i,-j}$ defined in Eq.~\eqref{ri-jdata}, it corresponds to the process where the noise $q_{-j}$ propagates in time to reach satellite $i$, and then subtracted by the clock noise $q_{i}$.
Such a term originated from the data $\tilde\eta^q_{-jk}$ propagates from $j$ to $i$ in the direction of time by applying ${\cal D}_{ij}$, where by definition, the data associated with the second node $i$ is $\tilde\eta^q_{ij}$.
It corresponds to the subfigure (b) shown in Fig.~\ref{fig4}.
Even though the process is visually similar to that given by subfigure (c), it gives a non-trivial contribution regarding the residual clock noise.

For the third variable $r_{-i,-j}$ defined in Eq.~\eqref{r-i-jdata}, it corresponds to the process where the noise $q_{-j}$ propagates in the opposite direction of time to the satellite $i$, and then subtracted by the clock noise $q_{-i}$.
Such a term originated from the data $\tilde\eta^q_{-jk}$ propagates from $j$ to $i$ in the direction of time, by applying ${\cal D}_{-ij}$, where by definition, the data associated with the second node $i$ is $\tilde\eta^q_{-ij}$.
It corresponds to the subfigure (d) shown in Fig.~\ref{fig4}.
It gives a non-trivial contribution regarding the residual clock noise.

The remaining subfigure (a) corresponds to the definition Eq.~\eqref{rrij}, originally proposed in~\cite{tdi-clock-2021}.
It is apparent that the four subfigures shown in Fig.~\ref{fig4} exhaust all possible trajectories that give rise to distinct contributions to the clock-noise residual.
Moreover, all these quantities can be eventually expressed in terms of the combination of the measurable quantity $r_{ij}$ that only involves the time-delay operation.
Nonetheless, we also note that the above definitions have not considered the second terms on the r.h.s. of Eqs.~\eqref{etaClockij} and~\eqref{eta-ij-FULL}.

By making use of the above results, we are now in the position to derive the clock-noise residual.
Specifically, by substituting Eqs.~\eqref{etaClockij} and~\eqref{eta-ij-FULL} into Eq.~\eqref{intereta}, we find
\begin{align}\label{clockresTOT}
	{\rm TDI}_{\rm geo}^q=\widetilde {\rm TDI}_{\rm geo}^q+\delta {\rm TDI}_{\rm geo}^q,
\end{align}
where
\begin{align}\label{clockres}
	&\widetilde {\rm TDI}_{\rm geo}^q=\nonumber\\
	&\!-\!\sum\limits_{k = 1}^n \left\{ \left( \prod\limits_{s = 1}^{k - 1} {\cal D}_{\pm i_{s - 1}i_s} \right)\tilde a _{\pm i_{k - 1}i_k}q_{\pm i_{k - 1}} \!-\! \left( \prod\limits_{s = 1}^{k - 1} {\cal D}_{\pm i'_{s - 1}i'_s} \right)\tilde a _{\pm i'_{k - 1}i'_k}q_{\pm i'_{k - 1}} \right\},
\end{align}
Furthermore, $\delta {\rm TDI}_{\rm geo}^q$ is a function of $r_{ij}$ due to the specific forms of the second terms on the r.h.s. of Eqs.~\eqref{etaClockij} and~\eqref{eta-ij-FULL}.
Its explicit form will be given below in Eq.~\eqref{comrtermDelta}.

To proceed, we aim to rewrite the r.h.s. of Eq.~\eqref{clockresTOT} regarding the clock-noise observables $r_{\pm i,\pm j}$.
This can be achieved by noticing the equality
\begin{align}\label{xicondition}
	\sum\limits_{k = 1}^n \left( \tilde a _{ \pm i_{k - 1}{i_k}}- \tilde a _{ \pm i'_{k - 1} i'_k} \right) = 0.
\end{align}
This is because Eq.~\eqref{dequalarm} is valid for arbitrary quantities with six distinct subscripts ${ij}$.
Therefore, one may replace $d_{ij}$ by $\tilde a _{ij}$ in Eq.~\eqref{dequalarm} to find Eq.~\eqref{xicondition}.
In light of this, the r.h.s. of Eq.~\eqref{clockres} can be put into the form 
\begin{align}
	&-\sum\limits_{k = 1}^n \left\{ \left( \prod\limits_{s = 1}^{k - 1} {\cal D}_{\pm i_{s - 1}i_s} \right)\tilde a _{\pm i_{k - 1}i_k}q_{\pm i_{k - 1}} - \left( \prod\limits_{s = 1}^{k - 1} {\cal D}_{\pm i'_{s - 1}i'_s} \right)\tilde a _{\pm i'_{k - 1}i'_k}q_{\pm i'_{k - 1}} \right\} \nonumber\\
	&+ \sum\limits_{k = 1}^n\left\{\tilde a _{\pm i_{k - 1}i_k}-\tilde a _{\pm i'_{k - 1}i'_k} \right\}q_{i_0}. \nonumber
\end{align}
For a given $k$, we have, by factoring out the coefficient $\tilde a_{\pm i_{k-1}i_k}$ and employing the same trick for the laser residual noise,
\begin{align}\label{clockkth}
	&\left( {\prod\limits_{s = 1}^{k - 1} {\cal D}_{ \pm i_{s - 1}{i_s}} } \right)q_{ \pm i_{k - 1}} - q_{i_0} \nonumber\\
	=& {\cal D}_{ \pm {i_0}{i_1}}{\cal D}_{ \pm {i_1}{i_2}}\cdots {\cal D}_{ \pm i_{k - 3}i_{k - 2}}{\cal D}_{ \pm i_{k - 2}i_{k - 1}}q_{ \pm i_{k - 1}}- q_{i_0}\notag\\
	=& {\cal D}_{ \pm {i_0}i_1\cdots i_{k - 3}i_{k - 2}}({\cal D}_{ \pm i_{k - 2}i_{k - 1}}q_{ \pm i_{k - 1}} - q_{ \pm i_{k - 2}})\nonumber\\
	+&  {\cal D}_{ \pm {i_0}i_1\cdots i_{k - 4}{i_{k - 3}}}({\cal D}_{ \pm {i_{k - 3}}{i_{k - 2}}}{q_{ \pm {i_{k - 2}}}} - {q_{ \pm {i_{k - 2}}}})\notag\\
	+& \cdots  + {\cal D}_{ \pm {i_0}{i_1}}({\cal D}_{ \pm {i_1}{i_2}}q_{\pm i_2}- q_{ \pm {i_1}}) + ({\cal D}_{ \pm {i_0}{i_1}}{q_ {\pm {i_1}}} - q_{{i_0}}),
\end{align}
which gives
\begin{align}\label{rkth}
	&r_{ \pm {i_0}{i_1}} + {\cal D}_{ \pm {i_0}{i_1}}r_{ \pm {i_1} \pm {i_2}} + \cdots  + {\cal D}_{ \pm {i_0}{i_1}}{\cal D}_{ \pm {i_1}{i_2}}\cdots {\cal D}_{ \pm i_{k - 4}i_{k - 3}}{\cal D}_{ \pm i_{k - 3}i_{k - 2}}r_{ \pm i_{k - 2} \pm i_{k - 1}}\nonumber\\ 
	&= \sum\limits_{m = 1}^{k - 1} {\left( {\prod\limits_{s = 1}^{m - 1} {\cal D}_{ \pm i_{s - 1}i_{s}} } \right)r_{ \pm {i_{m - 1}} \pm {i_m}}},
\end{align}
where one is reminded of the fact that all possible ways regarding clock noise are enumerated in Fig.~\ref{fig4} whose explicit forms are given in Eqs.~\eqref{rrij},~\eqref{r-ijdata},~\eqref{ri-jdata}, and~\eqref{r-i-jdata}.
By comparing the above derivations with those of the geometric TDI shown in Eq.~\eqref{interetaLASER}, there is a strong resemblance.
Specifically, the clock noise variables $r$ carry a similar role to $\eta$ for suppressing laser phase noise.

Putting all pieces together, we find
\begin{align}\label{comrterm}
	&\widetilde{\rm TDI}_{\rm geo}^q =\nonumber\\
	 -&\sum\limits_{k = 1}^n \sum\limits_{m = 1}^{k - 1} \left\{ \tilde a _{ \pm i_{k - 1}{i_k}}\left( \prod\limits_{s = 1}^{m - 1} {\cal D}_{ \pm i_{s - 1}i_{s}}  \right)r_{ \pm i_{m - 1} \pm i_m} \!-\! \tilde a _{ \pm i'_{k - 1}i'_k}\left( \prod\limits_{s = 1}^{m - 1} {\cal D}_{ \pm i'_{s - 1}i'_{s}} \right)r_{ \pm i'_{m - 1} \pm i'_m} \right\}.
\end{align}
Also, by similar arguments, we have
\begin{align}\label{comrtermDelta}
	\delta {\rm TDI}_{\rm geo}^q = \sum\limits_{k = 1}^n \left\{ \left( \prod\limits_{s = 1}^{k - 1} {\cal D}_{ \pm i_{s - 1}i_{s}}  \right)\delta \eta^q_{ \pm i_{k - 1} i_k} - \left( \prod\limits_{s = 1}^{k - 1} {\cal D}_{ \pm i_{s - 1}i_{s }}  \right)\delta \eta^q_{ \pm i'_{k - 1} i'_k} \right\},
\end{align}
where $\delta\eta^q_{ \pm ij}$ are defined by Eqs.~\eqref{delta1}.
It is noted that the specific form of $\delta {\rm TDI}^q$ is not particularly relevant because we can always redefine the intermediate variables $\eta^q_{\pm ij}$ by subtracting the observables $\delta \eta_{\pm ij}^q$, according to Eqs.~\eqref{etaClockij} and~\eqref{eta-ij-FULL}.

Finally, we arrive at the desired result with both the laser and clock noise suppressed
\begin{align}\label{suppreclockTOT}
	{\rm{TDI}}_{\rm{geo}}^{\rm{clock}} ={\rm{TDI}}_{\rm{geo}}^{\rm{laser}} -\widetilde{\rm{TDI}}_{\rm geo}^q -\delta {\rm TDI}_{\rm geo}^q,
\end{align}
where the r.h.s. of the equality are given, respectively, by Eqs.~\eqref{intereta},~\eqref{comrterm}, and~\eqref{comrtermDelta}. 

The noise residual given by Eq.~\eqref{suppreclockTOT} is the desired result for an arbitrary TDI solution derived using the geometric TDI approach.
Despite its tedious form consisting of double summation of monomials in time-shifted clock-noise observables, its explicit expression implies that it can be readily employed to eliminate the clock noise in geometric TDI solutions.
The next section will be devoted to discussing how to implement its evaluation as an algorithm and give a few explicit examples.

\section{Implementation}\label{section4}

In this section, we elaborate on how to implement the above results in terms of programming and computational expense.
Besides, we will further illustrate our findings with a few examples.

\subsection{Implementation of the clock-noise elimination scheme}\label{section4.1}

An arbitrary geometric TDI consists of two virtual equal-arm optical paths. 
The latter comprises successive links forming a closed trajectory joining discrete grids, as shown in Fig.~\ref{fig2}.
The horizontal direction is spatial, and the three possible values correspond to the indices of the three spacecraft. 
The vertical direction is temporal, and (positive) time evolution is upward.
As discussed in the previous section, all four possible scenarios are presented above in the subfigures (a-d) in Fig.~\ref{fig4} regarding the clock-noise residual.
In what follows, we specify how the clock-noise residual can be evaluated from a programming perspective.

Regarding the subfigure Fig.~\ref{fig4} (a), the residual clock noise up to the $k$th link $i_{k-1}i_k$ reads
\begin{align}\label{kthetaa}
	\prod\limits_{s = 1}^{k - 1} {\cal D}_{ \pm i_{s - 1}i_s}\tilde \eta^q _{i_{k - 1}i_k}  = {\cal D}_{ \pm {i_0}i_1 \cdots i_{k - 2}i_{k - 1}}\tilde \eta^q _{i_{k - 1}i_k}.
\end{align}
By substituting Eq.~\eqref{clockuniform}, one finds
\begin{align}\label{kthclocka}
	-{\cal D}_{ \pm i_0i_1 \cdots i_{k - 2}i_{k - 1}}{\tilde a _{i_{k - 1}i_k}}q_{i_{k - 1}}.
\end{align}
Subsequently, the residual clock noise can be rewritten in terms of the clock-noise observables by introducing a zeroth term
\begin{align}\label{kthra}
	&-\left( {\cal D}_{ \pm {i_0}i_1 \cdots i_{k - 2}{i_{k - 1}}}{\tilde a _{i_{k - 1}{i_k}}}{q_{{i_{k - 1}}}} - {\tilde a_{i_{k - 1}{i_k}}}{q_{{i_0}}} \right)\notag\\
	&= -{\tilde a_{{i_{k - 1}}{i_k}}}\left[ \begin{array}{l}
		{\cal D}_{ \pm {i_0}i_1 \cdots i_{k - 3}{i_{k - 2}}}\left( {{\cal D}_{{i_{k - 2}}{i_{k - 1}}}{q_{{i_{k - 1}}}} - {q_{{i_{k - 2}}}}} \right)+ \\
		{\cal D}_{ \pm {i_0}i_1\cdots i_{k - 4}{i_{k - 3}}}({\cal D}_{ \pm {i_{k - 3}}{i_{k - 2}}}{q_{{i_{k - 2}}}} - {q_{ \pm {i_{k - 3}}}})+\\
		\cdots  + {\cal D}_{ \pm {i_0}{i_1}}({\cal D}_{ \pm {i_1}{i_2}}{q_{ \pm {i_2}}} - {q_{ \pm {i_1}}}) + ({\cal D}_{ \pm {i_0}{i_1}}{q_{ \pm {i_1}}} - {q_{i_0}})
	\end{array} \right]\notag\\
	&= -\tilde a_{i_{k - 1}i_k}\left[ {\cal D}_{ \pm {i_0}i_1 \ldots i_{k - 3}i_{k - 2}}r_{i_{k - 2}i_{k - 1}} +  \ldots  + {\cal D}_{ \pm {i_0}{i_1}}r_{ \pm {i_1}{i_2}} + r_{ \pm {i_0} \pm {i_1}} \right]\notag\\
	&= -\tilde a_{i_{k - 1}i_k}\left[ {\cal D}_{ \pm {i_0}i_1 \ldots i_{k - 3}i_{k - 2}}r_{i_{k - 2}i_{k - 1}} + {\vartheta _{k - 2}} \right]\notag\\
	&= -\tilde a_{i_{k - 1}i_k}{\vartheta _{k - 1}},
\end{align}
where 
\begin{equation*}
	\vartheta _{j} \equiv \sum\limits_{m = 1}^{j} \left( \prod\limits_{s = 1}^{m - 1} {\cal D}_{ \pm i_{s - 1}i_s} \right)r_{ \pm i_{m - 1} \pm i_m} ,\nonumber
\end{equation*}
can be defined in a recursive fashion.
Specifically, we note that $\vartheta _{k-1}$ represents  the clock-noise suppression expression of the $k$-th link, which in turn can be obtained via $\vartheta _{k - 2} $ as given in Eq.~\eqref{kthra}.
By including all the contributions from these $\vartheta$ terms one finds the desired expression to eliminate the clock noise of the specific TDI combination.


For the subfigure, Fig.~\ref{fig4} (b), the residual clock noise up to the $k$th link $i_{k-1}i_k$ reads
\begin{align}\label{kthetab}
	\prod\limits_{s = 1}^{k - 1} {\cal D}_{\pm i_{s - 1}{i_s}} \tilde \eta^q _{ - i_{k - 1}i_k} =  - {\cal D}_{\pm {i_0}i_1 \ldots i_{k - 2}i_{k - 1}}{\cal D}_{ - i_{k - 1}i_k}\tilde \eta^q _{i_k i_{k - 1}}.
\end{align}
By substituting Eq.~\eqref{clockuniform}, we have
\begin{align}\label{kthclockb}
	-{\cal D}_{ \pm {i_0}i_1 \ldots i_{k - 2}i_{k - 1}}{\tilde a _{ - {i_{k - 1}}{i_k}}}{q_{ - {i_{k - 1}}}} =   {\cal D}_{ \pm {i_0}i_1 \ldots i_{k - 2}{i_{k - 1}}}{\cal D}_{ - {i_{k - 1}}{i_k}}\tilde a_{i_k i_{k - 1}}{q_k}.
\end{align}
Subsequently, the residual clock noise can be rewritten by introducing a zeroth term
\begin{align}\label{kthrb}
	&-\left({\cal D}_{ \pm {i_0}i_1 \ldots i_{k - 2}i_{k - 1}}\tilde a _{ - i_{k - 1}{i_k}}{q_{ -i_{k - 1}}} - \tilde a _{ - i_{k - 1}{i_k}}{q_{i_0}}\right) \notag\\
	&= -\tilde a _{ -i_{k - 1}{i_k}}\left[ \begin{array}{l}
		{\cal D}_{ \pm {i_0}i_1 \ldots i_{k - 3}i_{k - 2}}({\cal D}_{i_{k - 2}i_{k - 1}}{q_{ - i_{k - 1}}} - {q_{{i_{k - 2}}}}) +\\ {\cal D}_{ \pm {i_0}i_1\cdots i_{k - 4}i_{k - 3}}({\cal D}_{ \pm i_{k - 3}i_{k - 2}}{q_{i_{k - 2}}} - {q_{ \pm i_{k - 3}}}) + \\
		\cdots  + {\cal D}_{ \pm {i_0}{i_1}}({\cal D}_{ \pm {i_1}{i_2}}{q_{ \pm i_2}} - {q_{ \pm i_1}}) + ({\cal D}_{ \pm {i_0}{i_1}}{q_{ \pm i_1}} - {q_{i_0}})
	\end{array} \right]\notag\\
	&=-\tilde a _{ - i_{k - 1}{i_k}}\left[\begin{array}{l} 
		{\cal D}_{ \pm {i_0}i_1 \ldots i_{k - 3}i_{k - 2}}r_{i_{k - 2}, - i_{k - 1}} + {\cal D}_{ \pm {i_0}i_1\cdots i_{k - 4}i_{k - 3}}r_{ \pm i_{k - 3} \pm i_{k - 2}} +\\ \cdots  + {\cal D}_{ \pm {i_0}{i_1}}r_{ \pm {i_1} \pm {i_2}} + r_{ \pm {i_0} \pm {i_1}} 
	\end{array}\right]\notag\\
	&= -\tilde a _{ - i_{k - 1}{i_k}}\left[ {\cal D}_{ \pm {i_0}i_1 \ldots i_{k - 3}i_{k - 2}}\left( - {\cal D}_{i_{k - 2}i_{k - 1}}{\cal D}_{ -i_{k - 1}i_k}r_{i_k i_{k - 1}} + r_{i_{k - 2}i_{k - 1}} \right) + \vartheta _{k - 2} \right]\notag\\
	&= -\tilde a _{ -i_{k - 1}{i_k}}\left[ - {\cal D}_{ \pm {i_0}i_1 \ldots {i_{k-1}}{i_k}}r_{i_k i_{k - 1}} + {\cal D}_{ \pm {i_0}i_1 \ldots i_{k - 1}i_{k - 2}}r_{i_{k - 2}i_{k - 1}} + \vartheta _{k - 2} \right]\notag\\
	&= -\tilde a _{ -i_{k - 1}{i_k}} \vartheta _{k - 1} ,
\end{align}
Different from the case of Fig.~\ref{fig4} (a), the contribution from the link gives rise to two terms.

For the subfigure Fig.~\ref{fig4} (c), the residual noise up to the $k$th link reads
\begin{align}\label{kthetac}
	\prod\limits_{s = 1}^{k - 1} {\cal D}_{ \pm i_{s - 1}{i_s}}\tilde \eta^q _{i_{k - 1}{i_k}}  = {\cal D}_{ \pm {i_0}i_1 \cdots i_{k - 2}i_{k - 1}}\tilde \eta^q _{i_{k - 1}{i_k}}.
\end{align}
It gives
\begin{align}\label{kthclockc}
	-{\cal D}_{ \pm {i_0}i_1 \cdots i_{k - 2}i_{k - 1}}\tilde a _{i_{k - 1}{i_k}}{q_{i_{k - 1}}},
\end{align}
which can be further simplified by introducing the zeroth term
\begin{align}\label{kthrc}
	&-\left( {\cal D}_{ \pm i_0i_1\cdots i_{k - 2}i_{k - 1}}\tilde a _{i_{k - 1}i_k}q_{i_{k - 1}} - \tilde a _{i_{k - 1}i_k}q_{i_0} \right)\notag\\
	&= -\tilde a _{i_{k - 1}{i_k}}\left[ \begin{array}{l}
		{\cal D}_{ \pm {i_0}i_1 \cdots i_{k - 3}i_{k - 2}}({\cal D}_{ - i_{k - 2}i_{k - 1}}{q_{{i_{k - 1}}}} - {q_{ - {i_{k - 2}}}}) +\\ {\cal D}_{ \pm {i_0}i_1\cdots i_{k - 4}i_{k - 3}}({\cal D}_{ \pm i_{k - 3}i_{k - 2}}{q_{ - {i_{k - 2}}}} - {q_{ \pm {i_{k - 3}}}})\\
		+ \cdots  + {\cal D}_{ \pm {i_0}{i_1}}({\cal D}_{ \pm {i_1}{i_2}}{q_{ \pm {i_2}}} - {q_{ \pm {i_1}}}) + ({\cal D}_{ \pm {i_0}{i_1}}{q_{ \pm {i_1}}} - {q_{{i_0}}})
	\end{array} \right]\notag\\
	&= -\tilde a _{i_{k - 1}{i_k}}\left[ \begin{array}{l}
		{\cal D}_{ \pm {i_0}i_1 \cdots i_{k - 3}i_{k - 2}}r_{ - i_{k - 2},i_{k - 1}} + {\cal D}_{ \pm {i_0}i_1\cdots i_{k - 3}}r_{ \pm i_{k - 3} \pm i_{k - 2}} +\\ \cdots  + {\cal D}_{ \pm {i_0}{i_1}}r_{ \pm {i_1} \pm {i_2}} + r_{ \pm {i_0} \pm {i_1}} 
		\end{array} \right]\notag\\
	&= -\tilde a _{i_{k - 1}{i_k}}\left[ 0 + \vartheta _{k - 2}\right]\notag\\
	&= -\tilde a _{i_{k - 1}{i_k}}\vartheta _{k - 1}.
\end{align}
Locally, the link does not contribute to the residual clock noise in this case.

For the last subfigure Fig.~\ref{fig4}(d), we residual up to the $k$th link possesses the form
\begin{align}\label{kthetad}
	\prod\limits_{s = 1}^{k - 1} {\cal D}_{ \pm i_{k - 2}i_{k - 1}} \tilde \eta^q _{ -i_{k - 1}{i_k}} =  - {\cal D}_{ \pm {i_0}i_1 \ldots i_{k - 2}i_{k - 1}}{\cal D}_{ - i_{k - 1}{i_k}}\tilde \eta^q _{i_k i_{k - 1}}.
\end{align}
By substituting Eq.~\eqref{clockuniform}, we have
\begin{align}\label{kthclockd}
	-{\cal D}_{ \pm {i_0}i_1 \ldots i_{k - 2}i_{k - 1}}\tilde a _{ - i_{k - 1}{i_k}}q_{ - i_{k - 1}},
\end{align}
which subsequently leads to
\begin{align}\label{kthrd}
	&-\left( {\cal D}_{ \pm {i_0}i_1 \ldots i_{k - 2}i_{k - 1}}\tilde a _{ - i_{k - 1}{i_k}}q_{ -i_{k - 1}} - \tilde a _{ - i_{k - 1}{i_k}}q_{i_0} \right)\notag\\
	&=-\tilde a _{ - i_{k - 1}{i_k}}\left[ \begin{array}{l}
		{\cal D}_{ \pm {i_0}i_1 \cdots i_{k - 3}i_{k - 2}}({\cal D}_{ - i_{k - 2}i_{k - 1}}q_{ - i_{k - 1}} - q_{ - i_{k - 2}}) +\\ {\cal D}_{ \pm {i_0}i_1\cdots i_{k - 4}i_{k - 3}}({\cal D}_{ \pm i_{k - 3}i_{k - 2}}q_{ -i_{k - 2}} -q_{ \pm i_{k - 3}})
		+\\ \cdots \! +\! {\cal D}_{ \pm {i_0}{i_1}}({\cal D}_{ \pm {i_1}{i_2}}q_{ \pm {i_2}} - q_{ \pm {i_1}}) \!+ \!({\cal D}_{ \pm {i_0}{i_1}}q_{ \pm {i_1}} - q_{i_0})
	\end{array} \right]\notag\\
	&= -\tilde a _{ - i_{k - 1}{i_k}}\left[ \begin{array}{l}
		{\cal D}_{ \pm {i_0}i_1 \cdots i_{k - 3}i_{k - 2}}r_{ - i_{k - 2}, - i_{k - 1}} + {\cal D}_{ \pm {i_0}i_1\cdots i_{k - 4}i_{k - 3}}r_{ \pm i_{k - 3} \pm i_{k - 2}} +\\ \cdots  + {\cal D}_{ \pm {i_0}{i_1}}r_{ \pm {i_1} \pm {i_2}} + r_{ \pm {i_0} \pm {i_1}} 
		\end{array} \right]\notag\\
	&= -\tilde a _{- i_{k - 1}{i_k}}\left[ \begin{array}{l}
		- {\cal D}_{ \pm {i_0}i_1 \ldots i_{k - 3}i_{k - 2}}{\cal D}_{ - i_{k - 2}i_{k - 1}}{\cal D}_{ - i_{k - 1}{i_k}}r_{i_k i_{k - 1}}+ \\{\cal D}_{ \pm {i_0}i_1\cdots i_{k - 2}i_{k - 3}}r_{ \pm i_{k - 3} \pm i_{k - 2}} \!+\! \cdots \! +\! {\cal D}_{ \pm {i_0}{i_1}}r_{ \pm {i_1} \pm {i_2}} + r_{ \pm {i_0} \pm {i_1}}
		\end{array} \right]\notag\\
	&= -\tilde a _{ - i_{k - 1}{i_k}}\left[ - {\cal D}_{ \pm {i_0}i_1 \ldots i_{k-1}{i_k}}r_{i_k i_{k - 1}} + \vartheta _{k - 2} \right] \notag\\
	&= -\tilde a _{ - i_{k - 1}{i_k}} \vartheta _{k - 1}.
\end{align}

The above results are summarized in Tab.~\ref{suppressclock}.
Following the discussions in the last section and the definitions of $r_{\pm i,\pm j}$, the results derived in Eqs.~\eqref{kthra},~\eqref{kthrb},~\eqref{kthrc}, and~\eqref{kthrd} can be readily utilized to evaluate the residual clock noise in a geometric TDI solution.
The contributions from a specific link fall in one of the four possibilities.
They are essentially comprised of the clock-noise observables subject to successive time-shift operations, expressed explicitly in terms of products of time-shift operators.
Based on the above notation, the contributions from the remaining $k-2$ links, namely, $\vartheta _{k - 2}$, can be obtained straightforwardly.
From a programming perspective, these terms can be evaluated in an exhaustive manner beforehand, independent of the specific geometrical TDI solution.
Subsequently, clock noise subtraction can be readily performed with efficiency.

\begin{table}
	\centering
	\caption{A summary of the results obtained in Eqs.~\eqref{kthra},~\eqref{kthrb},~\eqref{kthrc}, and~\eqref{kthrd}.
		The time-shifted intermediate variable $\tilde \eta_{\pm ij}$, residual clock noise, and the latter expressed in terms of the clock-noise observables $r_{ij}$.}
	\newcommand{\tabincell}[2]{\begin{tabular}{@{}#1@{}}#2\end{tabular}}
	\renewcommand\arraystretch{1.5}
	\resizebox{\textwidth}{!}{
		\begin{tabular}{|c|c|c|c|}
			\hline
			\hline
			type & \makecell[c]{time-shifted intermediate variable\\ for the $k$th link} & {residual clock noise}& form in clock-noise observables\\
			\hline
			$a$&${\cal D}_{ \pm {i_0}i_1 \cdots i_{k - 2}i_{k - 1}}\tilde \eta^q _{i_{k - 1}i_k}$&$-{\cal D}_{ \pm i_0i_1 \cdots i_{k - 2}i_{k - 1}}{\tilde a _{i_{k - 1}i_k}}q_{i_{k - 1}}$&$-\tilde a _{i_{k - 1}i_k}\left[ {\cal D}_{ \pm {i_0}i_1 \ldots i_{k - 3}i_{k - 2}}r_{i_{k - 2}i_{k - 1}} + {\vartheta _{k - 2}} \right]$\\
			\hline
			$b$&$- {\cal D}_{\pm {i_0}i_1 \ldots i_{k - 2}i_{k - 1}}{\cal D}_{ - i_{k - 1}i_k}\tilde\eta^q _{i_k i_{k - 1}}$&${\cal D}_{ \pm {i_0}i_1 \ldots i_{k - 2} i_{k - 1}}{\cal D}_{ - {i_{k - 1}}{i_k}}{\tilde a _{i_k i_{k - 1}}}{q_k}$&$-\tilde a _{ -i_{k - 1}{i_k}}\left[ - {\cal D}_{ \pm {i_0}i_1 \ldots i_{k - 1}{i_k}}r_{i_k i_{k - 1}} + {\cal D}_{ \pm {i_0}i_1 \ldots i_{k - 3}i_{k - 2}}r_{i_{k - 2}i_{k - 1}} + \vartheta _{k - 2} \right]$ \\
			\hline
			$c$&${\cal D}_{ \pm {i_0}i_1 \cdots i_{k - 2}i_{k - 1}}\tilde\eta^q _{i_{k - 1}{i_k}}$&$-{\cal D}_{ \pm {i_0}i_1 \cdots i_{k - 2}i_{k - 1}}\tilde a _{i_{k - 1}{i_k}}{q_{i_{k - 1}}}$&$-\tilde a _{i_{k - 1}{i_k}}\left[ 0 + \vartheta _{k - 2}\right]$ \\
			\hline
			$d$&$- {\cal D}_{ \pm {i_0}i_1 \ldots i_{k - 2}i_{k - 1}}{\cal D}_{ - i_{k - 1}{i_k}}\tilde\eta^q _{i_k i_{k - 1}}$&$-{\cal D}_{ \pm {i_0}i_1 \ldots i_{k - 2}i_{k - 1}}\tilde a _{ - i_{k - 1}{i_k}}q_{ - i_{k - 1}}$&$-\tilde a _{ - i_{k - 1}{i_k}}\left[ - {\cal D}_{ \pm {i_0}i_1 \ldots i_{k - 1}{i_k}}r_{i_k i_{k - 1}} + \vartheta _{k - 2} \right]$ \\
			\hline
			\hline
		\end{tabular}
	}
	\label{suppressclock}
\end{table}

\subsection{A few examples}\label{section4.2}

Now, we are in the position to illustrate our results with a few examples.

\begin{figure}[H]
	\centering
	\includegraphics[width=0.3\textwidth]{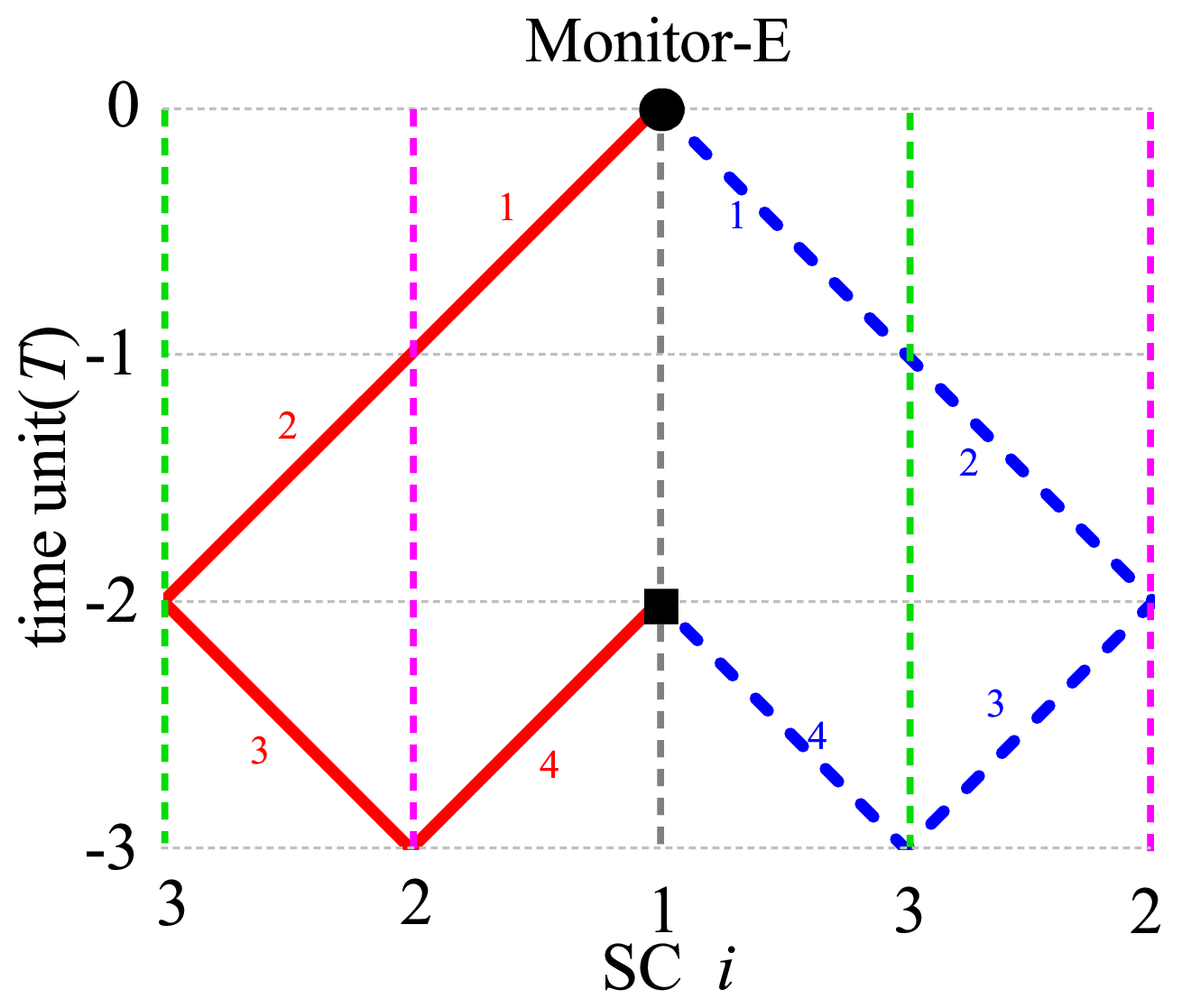}\hspace{30pt}
	\includegraphics[width=0.35\textwidth]{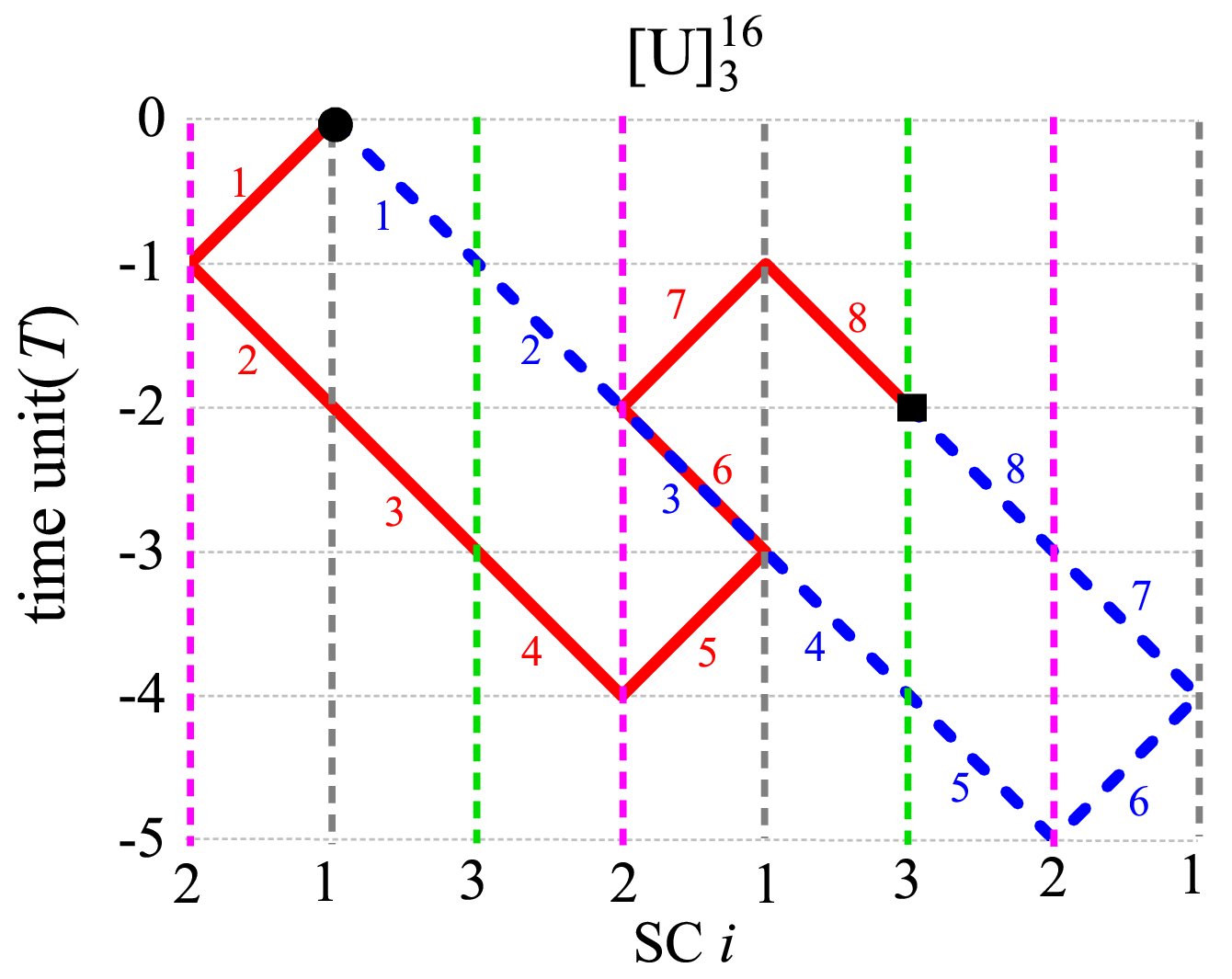}
	\caption{\label{fig5}
		The spacetime diagrams of the first-generation eight-link Monitor combination $E_1$ (left) and the sixteen-link Relay combination $\left[ {\rm{U}} \right]_3^{16}$ (right).
		The red-solid and blue-dashed line segments correspond to the two synthesized routes, which might contain both forward and backward propagation in time.
		The black squares and black dots indicate the initial and terminal grids of the two routes, and the reference time $t=0$ is attributed to the terminal SC 1.}
\end{figure}

As a first example, we consider the first-generation geometric TDI Monitor-E combination, whose space-time diagram is shown in the left panel of Fig.~\ref{fig5}.
In Tabs.~\ref{suppressclockmonred} and~\ref{suppressclockmonblue}, we show, for individual links, the time-shifted intermediate variable $\eta$, clock noise, and the latter presented in clock-noise observables.
It is readily verified that the introduced zeroth terms in Eqs.~\eqref{kthra},~\eqref{kthrb},~\eqref{kthrc}, and~\eqref{kthrd}, precisely cancel each other.
To be specific, all those zeroth terms are proportional to $q_1$, and the sum of the contributions corresponding to the red-solid and blue-dashed line segments give
\begin{align}\label{ximonq10}
	\left( \tilde a _{12} +\tilde a _{23} + \tilde a_{32} - \tilde a _{12} \right){q_1} - \left(  \tilde a _{13}+ \tilde a _{32}+ \tilde a _{23} -\tilde a _{13} \right){q_1}=0 ,
\end{align}
in accordance with Eq.~\eqref{xicondition}.

\begin{table}
	\centering
	\caption{The relevant quantities for evaluating the residual clock noise corresponding to the four links in red solid line segments of the geometric TDI solution Monitor-E combination shown in the left panel of Fig~\ref{fig5}.}
	\newcommand{\tabincell}[2]{\resizebox{\textwidth}{!}{%
\begin{tabular}{@{}#1@{}}#2\end{tabular}}
	\renewcommand\arraystretch{1.5}
	\begin{tabular}{|c|c|c|c|c|}
		\hline
		\hline
		link & {intermediate variable}&clock noise & form in clock-noise observables & type \\
		\hline
		$1$&$\tilde \eta^q _{12}$&$-\tilde a _{12}{q_1} + \tilde a_{12}{q_1}$&$0$&\\
		\hline
		$2$&${\cal D}_{12}\tilde \eta^q _{23}$&$-\tilde a _{23}{\cal D}_{12}{q_2} + \tilde a _{23}{q_1}$&$ -\tilde a _{23}r_{12}$&$a$ \\
		\hline
		$3$&${\cal D}_{12}{\cal D} _{23}\tilde \eta^q_{32}$&$-\tilde a_{32}{\cal D}_{12}{\cal D} _{23}{q_3} + \tilde a _{32}{q_1}$&$-\tilde a _{32}({\cal D}_{12}r_{23}+r_{12})$&$a$ \\
		\hline
		$4$&$-{\cal D}_{12}{\cal D} _{23}{\cal D} _{32}{\cal D} _{-21}\tilde \eta^q_{12}$&$\tilde a _{12}{\cal D}_{12}{\cal D} _{23}{\cal D} _{32}{\cal D} _{-21}{q_1} - \tilde a _{12}{q_1}$&$\tilde a _{12}(-{\cal D}_{12}{\cal D} _{23}{\cal D} _{32}{\cal D} _{-21}r_{12}+{\cal D}_{12}{\cal D}_{23}r_{32}+{\cal D}_{12}r_{23}+r_{12})$&$b$ \\
		\hline
		\hline
	\end{tabular}}
	\label{suppressclockmonred}
\end{table}

\begin{table}
	\centering
	\caption{The relevant quantities for evaluating the residual clock noise corresponding to the four links in blue dashed line segments of the geometric TDI solution Monitor-E combination shown in the left panel of Fig~\ref{fig5}.}
	\newcommand{\tabincell}[2]{\resizebox{\textwidth}{!}{%
\begin{tabular}{@{}#1@{}}#2\end{tabular}}
	\renewcommand\arraystretch{1.5}
	\begin{tabular}{|c|c|c|c|c|}
		\hline
		\hline
		link & {intermediate variable}&clock noise&\makecell[c]{form in clock-noise observables}&type \\
		\hline
		$1$&$\tilde \eta^q _{13}$&$-\tilde a _{13}{q_1} + \tilde a _{13}{q_1}$&$0$&\\
		\hline
		$2$&${\cal D}_{13}\tilde \eta^q _{32}$&$-\tilde a _{32}{\cal D}_{13}{q_3} + \tilde a _{32}{q_1}$&$-\tilde a _{32}r_{13}$&$a$ \\
		\hline
		$3$&${\cal D}_{13}{\cal D} _{32}\tilde \eta^q _{23}$&$-\tilde a _{23}{\cal D}_{13}{\cal D} _{32}{q_2} + \tilde a _{23}{q_1}$&$-\tilde a _{23}({\cal D}_{13}r_{32}+r_{13})$&$a$ \\
		\hline
		$4$&$-{\cal D}_{13}{\cal D} _{32}{\cal D} _{23}{\cal D}_{-31}\tilde \eta^q _{13}$&$\tilde a _{13}{\cal D}_{13}{\cal D} _{32}{\cal D} _{23}{\cal D}_{-31}{q_1} -\tilde a _{13}{q_1}$&$\tilde a _{13}(-{\cal D}_{13}{\cal D} _{32}{\cal D} _{23}{\cal D}_{-31}r_{13}+{\cal D}_{13}{\cal D} _{32}r_{23}+{\cal D}_{13}r_{32}+r_{13})$&$b$ \\
		\hline
		\hline
	\end{tabular}}
	\label{suppressclockmonblue}
\end{table}

Our second example is the second-generation Michelson TDI combination $[X]_1^{16}$. 
In Tabs.~\ref{suppressclockredx2com} and~\ref{suppressclockbluex2com}, we show the relevant quantities for clock-noise reduction scheme for individual links.
Again, all those zeroth terms are proportional to $q_1$, and the sum of the contributions corresponding to the red-solid and blue-dashed line segments readily vanishes in accordance with Eq.~\eqref{xicondition}, namely, 
\begin{align}\label{xx2q10}
	&\left( \tilde a _{12} +\tilde a _{21} + \tilde a_{13} + \tilde a _{31}+ \tilde a _{13} + \tilde a _{31} + \tilde a _{12} + \tilde a _{21} \right){q_1} \nonumber\\
	&- \left( \tilde a _{13}+ \tilde a _{31}+ \tilde a _{12} + \tilde a _{21} + \tilde a _{12}+ \tilde a _{21} +\tilde a _{13} +\tilde a _{31} \right){q_1}=0 .
\end{align}

\begin{table}
	\centering
	\caption{The relevant quantities for evaluating the residual clock noise corresponding to the eight links in red solid line segments of the geometric TDI solution $[X]_1^{16}$ shown in the right panel of Fig~\ref{fig2}.}
	\newcommand{\tabincell}[2]{\resizebox{\textwidth}{!}{%
\begin{tabular}{@{}#1@{}}#2\end{tabular}}
	\renewcommand\arraystretch{1.5}
	\begin{tabular}{|c|c|c|c|c|}
		\hline
		\hline
		link & {intermediate variable}&clock noise&\makecell[c]{form in clock-noise observables}&type \\
		\hline
		$1$&$\tilde \eta^q _{12}$&$-\tilde a _{12}{q_1} + \tilde a_{12}{q_1}$&$0$&\\
		\hline
		$2$&${\cal D}_{12}\tilde \eta^q _{21}$&$-\tilde a _{21}{\cal D}_{12}{q_2} + \tilde a _{21}{q_1}$&$ -\tilde a _{21}r_{12}$&$a$ \\
		\hline
		$3$&${\cal D}_{12}{\cal D} _{21}\tilde \eta^q_{13}$&$-\tilde a _{13}{\cal D}_{12}{\cal D} _{21}{q_1} + \tilde a _{13}{q_1}$&$-\tilde a _{13}({\cal D}_{12}r_{21}+r_{12})$&$a$ \\
		\hline
		$4$&${\cal D}_{12}{\cal D} _{21}{\cal D} _{13}\tilde \eta^q_{31}$&$-\tilde a _{31}{\cal D}_{12}{\cal D} _{21}{\cal D} _{13}{q_3} + \tilde a _{31}{q_1}$&$-\tilde a _{31}({\cal D}_{12}{\cal D} _{21}r_{13}+{\cal D}_{12}r_{21}+r_{12})$&$a$ \\
		\hline
		$5$&${\cal D}_{12}{\cal D} _{21}{\cal D} _{13}{\cal D}_{31}\tilde \eta^q_{13}$&$-\tilde a _{13}{\cal D}_{12}{\cal D} _{21}{\cal D} _{13}{\cal D}_{31}{q_1}+ \tilde a _{13}{q_1}$&$\makecell[c]{ -\tilde a _{13}\left( \begin{array}{l}{\cal D}_{12}{\cal D} _{21}{\cal D} _{13}r_{31}\\+{\cal D}_{12}{\cal D} _{21}r_{13}+{\cal D}_{12}r_{21}+r_{12}\end{array} \right)}$&$a$ \\
		\hline
		$6$&${\cal D}_{12}{\cal D} _{21}{\cal D} _{13}{\cal D}_{31}{\cal D}_{13}\tilde \eta^q _{31}$&\makecell[c]{$-\tilde a _{31}{\cal D}_{12}{\cal D} _{21}{\cal D} _{13}{\cal D} _{31}{\cal D}_{13}{q_3}$\\+$\tilde a _{31}{q_1}$}&\makecell[c]{$ -\tilde a _{31}\left( \begin{array}{l}
				{\cal D}_{12}{\cal D} _{21}{\cal D} _{13}{\cal D}_{31}r_{13}\\
				+{\cal D}_{12}{\cal D} _{21}{\cal D} _{13}r_{31}\\
				+{\cal D}_{12}{\cal D} _{21}r_{13}+{\cal D}_{12}r_{21}+r_{12}
			\end{array} \right)$}&$a$ \\
		\hline
		$7$&${\cal D}_{12}{\cal D} _{21}{\cal D} _{13}{\cal D}_{31}{\cal D}_{13}{\cal D}_{31}\tilde \eta^q _{12}$&\makecell[c]{$-\tilde a _{12}{\cal D}_{12}{\cal D} _{21}{\cal D} _{13}{\cal D}_{31}{\cal D}_{13}{\cal D}_{31}q_1$\\ +$\tilde a _{12}{q_1}$}&\makecell[c]{$ -\tilde a _{12}\left( \begin{array}{l}
				{\cal D}_{12}{\cal D} _{21}{\cal D} _{13}{\cal D}_{31}{\cal D}_{13}r_{31}\\
				+{\cal D}_{12}{\cal D} _{21}{\cal D} _{13}{\cal D}_{31}r_{13}\\
				+{\cal D}_{12}{\cal D} _{21}{\cal D} _{13}r_{31}\\
				+{\cal D}_{12}{\cal D} _{21}r_{13}+{\cal D}_{12}r_{21}+r_{12}
			\end{array} \right)$}&$a$ \\
		\hline
		$8$&${\cal D}_{12}{\cal D} _{21}{\cal D} _{13}{\cal D}_{31}{\cal D}_{13}{\cal D}_{31}{\cal D}_{12}\tilde \eta^q _{21}$&$\makecell[c]{-\tilde a _{21}{\cal D}_{12}{\cal D} _{21}{\cal D} _{13}{\cal D}_{31}{\cal D}_{13}{\cal D}_{31}{\cal D}_{12}{q_2} \\+\tilde a _{21}{q_1}}$&\makecell[c]{$ -\tilde a _{21}\left( \begin{array}{l}
				{\cal D}_{12}{\cal D} _{21}{\cal D} _{13}{\cal D}_{31}{\cal D}_{13}{\cal D}_{31}r_{12}\\
				+{\cal D}_{12}{\cal D} _{21}{\cal D} _{13}{\cal D}_{31}{\cal D}_{13}r_{31}\\
				+{\cal D}_{12}{\cal D} _{21}{\cal D} _{13}{\cal D}_{31}r_{13}\\
				+{\cal D}_{12}{\cal D} _{21}{\cal D} _{13}r_{31}\\
				+{\cal D}_{12}{\cal D} _{21}r_{13}+{\cal D}_{12}r_{21}+r_{12}
			\end{array} \right)$}&$a$\\
		\hline
		\hline
	\end{tabular}}
	\label{suppressclockredx2com}
\end{table}

\begin{table}
	\centering
	\caption{The relevant quantities for evaluating the residual clock noise corresponding to the eight links in blue dashed line segments of the geometric TDI solution $[X]_1^{16}$ shown in the right panel of Fig~\ref{fig2}.}
	\newcommand{\tabincell}[2]{\resizebox{\textwidth}{!}{%
\begin{tabular}{@{}#1@{}}#2\end{tabular}}
	\renewcommand\arraystretch{1.5}
	\begin{tabular}{|c|c|c|c|c|}
		\hline
		\hline
		link& {intermediate variable}&clock noise&\makecell[c]{form in clock-noise observables}&type \\
		\hline
		$1$&$\tilde \eta^q _{13}$&$-\tilde a _{13}{q_1} + \tilde a _{13}{q_1}$&$0$&\\
		\hline
		$2$&${\cal D}_{13}\tilde \eta^q _{31}$&$-\tilde a _{31}{\cal D}_{13}{q_3} + \tilde a _{31}{q_1}$&$-\tilde a _{31}r_{13}$&$a$ \\
		\hline
		$3$&${\cal D}_{13}{\cal D}_{31}\tilde \eta^q _{12}$&$-\tilde a _{12}{\cal D}_{13}{\cal D}_{31}{q_1} + \tilde a _{12}{q_1}$&$-\tilde a _{12}({\cal D}_{13}r_{31}+r_{13})$&$a$ \\
		\hline
		$4$&${\cal D}_{13}{\cal D}_{31}{\cal D}_{12}\tilde \eta^q _{21}$&$-\tilde a _{21}{\cal D}_{13}{\cal D}_{31}{\cal D}_{12}{q_2} + \tilde a _{21}{q_1}$&$-\tilde a _{21}({\cal D}_{13}{\cal D}_{31}r_{12}+{\cal D}_{13}r_{31}+r_{13})$&$a$ \\
		\hline
		$5$&${\cal D}_{13}{\cal D}_{31}{\cal D}_{12}{\cal D}_{21}\tilde \eta^q _{12}$&$-\tilde a _{12}{\cal D}_{13}{\cal D}_{31}{\cal D}_{12}{\cal D}_{21}{q_1} +\tilde a _{12}{q_1}$&\makecell[c]{$-\tilde a _{12}\left( \begin{array}{l}{\cal D}_{13}{\cal D}_{31}{\cal D}_{12}r_{21}\\+{\cal D}_{13}{\cal D}_{31}r_{12}+{\cal D}_{13}r_{31}+r_{13}\end{array} \right)$}&$a$ \\
		\hline
		$6$&${\cal D}_{13}{\cal D}_{31}{\cal D}_{12}{\cal D}_{21}{\cal D}_{12}\tilde \eta^q _{21}$&$\makecell[c]{-\tilde a _{21}{\cal D}_{13}{\cal D}_{31}{\cal D}_{12}{\cal D}_{21}{\cal D}_{12}{q_2} \\+ \tilde a _{21}{q_1}}$&$\makecell[c]{-\tilde a _{21}\left( \begin{array}{l}
				{\cal D}_{13}{\cal D}_{31}{\cal D}_{12}{\cal D}_{21}r_{12}\\
				+{\cal D}_{13}{\cal D}_{31}{\cal D}_{12}r_{21}\\+{\cal D}_{13}{\cal D}_{31}r_{12}+{\cal D}_{13}r_{31}+r_{13}\end{array} \right)}$ &$a$\\
		\hline
		$7$&${\cal D}_{13}{\cal D}_{31}{\cal D}_{12}{\cal D}_{21}{\cal D}_{12}{\cal D}_{21}\tilde \eta^q_{13}$&$\makecell[c]{-\tilde a _{13}{\cal D}_{13}{\cal D}_{31}{\cal D}_{12}{\cal D}_{21}{\cal D}_{12}{\cal D}_{21}{q_1} \\+ \tilde a _{13}{q_1}}$&$\makecell[c]{-\tilde a _{13}\left( \begin{array}{l}
				{\cal D}_{13}{\cal D}_{31}{\cal D}_{12}{\cal D}_{21}{\cal D}_{12}r_{21}\\
				+{\cal D}_{13}{\cal D}_{31}{\cal D}_{12}{\cal D}_{21}r_{12}\\
				+{\cal D}_{13}{\cal D}_{31}{\cal D}_{12}r_{21}\\+{\cal D}_{13}{\cal D}_{31}r_{12}+{\cal D}_{13}r_{31}+r_{13}\end{array} \right)}$&$a$ \\
		\hline
		$8$&${\cal D}_{13}{\cal D}_{31}{\cal D}_{12}{\cal D}_{21}{\cal D}_{12}{\cal D}_{21}{\cal D}_{13}\tilde \eta^q_{31}$&$\makecell[c]{-\tilde a _{31}{\cal D}_{13}{\cal D}_{31}{\cal D}_{12}{\cal D}_{21}{\cal D}_{12}{\cal D}_{21}{\cal D}_{13}{q_3} \\+ \tilde a _{31}{q_1}}$&$\makecell[c]{-\tilde a _{31}\left( \begin{array}{l}
				{\cal D}_{13}{\cal D}_{31}{\cal D}_{12}{\cal D}_{21}{\cal D}_{12}{\cal D}_{21}r_{13}\\
				+{\cal D}_{13}{\cal D}_{31}{\cal D}_{12}{\cal D}_{21}{\cal D}_{12}r_{21}\\
				+{\cal D}_{13}{\cal D}_{31}{\cal D}_{12}{\cal D}_{21}r_{12}\\
				+{\cal D}_{13}{\cal D}_{31}{\cal D}_{12}r_{21}\\+{\cal D}_{13}{\cal D}_{31}r_{12}+{\cal D}_{13}r_{31}+r_{13}\end{array} \right)}$&$a$ \\
		\hline
		\hline
	\end{tabular}}
	\label{suppressclockbluex2com}
\end{table}


\begin{table}
	\centering
	\caption{The relevant quantities for evaluating the residual clock noise corresponding to the eight links in red solid line segments of the geometric TDI solution $\left[ {\rm{U}} \right]_3^{16}$ shown in the right panel of Fig~\ref{fig5}.}
	\newcommand{\tabincell}[2]{\begin{tabular}{@{}#1@{}}#2\end{tabular}}
	\renewcommand\arraystretch{1.5}
	\resizebox{\textwidth}{!}{
		\begin{tabular}{|c|c|c|c|c|}
			\hline
			\hline
			link & {intermediate variable}&clock noise&\makecell[c]{form in clock-noise observables}&type \\
			\hline
			$1$&$\tilde \eta^q _{12}$&$-\tilde a _{12}{q_1} + \tilde a_{12}{q_1}$&$0$&\\
			\hline
			$2$&${\cal D}_{12}\tilde \eta^q _{21}$&$-\tilde a _{21}{\cal D}_{12}{q_2} + \tilde a _{21}{q_1}$&$ -\tilde a _{21}r_{12}$&$a$ \\
			\hline
			$3$&${\cal D}_{12}{\cal D} _{21}\tilde \eta^q_{13}$&$-\tilde a _{13}{\cal D}_{12}{\cal D} _{21}{q_1} + \tilde a _{13}{q_1}$&$-\tilde a _{13}({\cal D}_{12}r_{21}+r_{12})$&$a$ \\
			\hline
			$4$&${\cal D}_{12}{\cal D} _{21}{\cal D} _{13}\tilde \eta^q_{32}$&$-\tilde a _{32}{\cal D}_{12}{\cal D} _{21}{\cal D} _{13}{q_3} + \tilde a _{32}{q_1}$&$-\tilde a _{32}({\cal D}_{12}{\cal D} _{21}r_{13}+{\cal D}_{12}r_{21}+r_{12})$&$a$ \\
			\hline
			$5$&$-{\cal D}_{12}{\cal D} _{21}{\cal D} _{13}{\cal D} _{32}{\cal D}_{-21}\tilde \eta^q_{12}$&$\makecell[c]{\tilde a _{12}{\cal D}_{12}{\cal D} _{21}{\cal D} _{13}{\cal D} _{32}{\cal D}_{-21}{q_1}\\ - \tilde a _{12}{q_1}}$&\makecell[c]{$ \tilde a _{12}\left( \begin{array}{l}
					-{\cal D}_{12}{\cal D} _{21}{\cal D} _{13}{\cal D} _{32}{\cal D}_{-21}r_{12}\\
					+{\cal D}_{12}{\cal D} _{21}{\cal D}_{13}r_{32}\\
					+{\cal D}_{12}{\cal D} _{21}r_{13}+{\cal D}_{12}r_{21}+r_{12}
				\end{array} \right)$}&$b$ \\
			\hline
			$6$&$-{\cal D}_{12}{\cal D} _{21}{\cal D} _{13}{\cal D} _{32}{\cal D}_{-21}{\cal D}_{-12}\tilde \eta^q _{21}$&$\makecell[c]{\tilde a _{21}{\cal D}_{12}{\cal D} _{21}{\cal D} _{13}{\cal D} _{32}{\cal D}_{-21}{\cal D}_{-12}{q_2}\\ - \tilde a _{21}{q_1}}$&\makecell[c]{$ \tilde a _{21}\left( \begin{array}{l}
					-{\cal D}_{12}{\cal D} _{21}{\cal D} _{13}{\cal D} _{32}{\cal D}_{-21}{\cal D}_{-12}r_{21}\\
					-{\cal D}_{12}{\cal D} _{21}{\cal D} _{13}{\cal D} _{32}{\cal D}_{-21}r_{12}\\
					+{\cal D}_{12}{\cal D} _{21}{\cal D}_{13}r_{32}\\
					+{\cal D}_{12}{\cal D} _{21}r_{13}+{\cal D}_{12}r_{21}+r_{12}
				\end{array} \right)$}&$d$ \\
			\hline
			$7$&$-{\cal D}_{12}{\cal D} _{21}{\cal D} _{13}{\cal D} _{32}{\cal D}_{-21}{\cal D}_{-12}{\cal D}_{-21}\tilde \eta^q _{12}$&$\makecell[c]{\tilde a _{12}{\cal D}_{12}{\cal D} _{21}{\cal D} _{13}{\cal D} _{32}{\cal D}_{-21}{\cal D}_{-12}{\cal D}_{-21}{q_1}\\ - \tilde a _{12}{q_1}}$&\makecell[c]{$ \tilde a _{12}\left( \begin{array}{l}
					-{\cal D}_{12}{\cal D} _{21}{\cal D} _{13}{\cal D} _{32}{\cal D}_{-21}{\cal D}_{-12}{\cal D}_{-21}r_{12}\\
					-{\cal D}_{12}{\cal D} _{21}{\cal D} _{13}{\cal D} _{32}{\cal D}_{-21}{\cal D}_{-12}r_{21}\\
					-{\cal D}_{12}{\cal D} _{21}{\cal D} _{13}{\cal D} _{32}{\cal D}_{-21}r_{12}\\
					+{\cal D}_{12}{\cal D} _{21}{\cal D}_{13}r_{32}\\
					+{\cal D}_{12}{\cal D} _{21}r_{13}+{\cal D}_{12}r_{21}+r_{12}
				\end{array} \right)$}&$d$ \\
			\hline
			$8$&$-{\cal D}_{12}{\cal D} _{21}{\cal D} _{13}{\cal D} _{32}{\cal D}_{-21}{\cal D}_{-12}{\cal D}_{-21}\tilde \eta^q_{13}$&$\makecell[c]{\tilde a _{13}{\cal D}_{12}{\cal D} _{21}{\cal D} _{13}{\cal D} _{32}{\cal D}_{-21}{\cal D}_{-12}{\cal D}_{-21}{q_1} \\+\tilde a _{13}{q_1}}$&\makecell[c]{$ \tilde a _{13}\left( \begin{array}{l}
					-{\cal D}_{12}{\cal D} _{21}{\cal D} _{13}{\cal D} _{32}{\cal D}_{-21}{\cal D}_{-12}{\cal D}_{-21}r_{12}\\
					-{\cal D}_{12}{\cal D} _{21}{\cal D} _{13}{\cal D} _{32}{\cal D}_{-21}{\cal D}_{-12}r_{21}\\
					-{\cal D}_{12}{\cal D} _{21}{\cal D} _{13}{\cal D} _{32}{\cal D}_{-21}r_{12}\\
					+{\cal D}_{12}{\cal D} _{21}{\cal D}_{13}r_{32}\\
					+{\cal D}_{12}{\cal D} _{21}r_{13}+{\cal D}_{12}r_{21}+r_{12}
				\end{array} \right)$}&$c$ \\
			\hline
			\hline
		\end{tabular}
	}
	\label{suppressclockred}
\end{table}

\begin{table}
	\centering
	\caption{The relevant quantities for evaluating the residual clock noise corresponding to the eight links in blue dashed line segments of the geometric TDI solution $\left[ {\rm{U}} \right]_3^{16}$ shown in the right panel of Fig~\ref{fig5}.}
	\newcommand{\tabincell}[2]{\begin{tabular}{@{}#1@{}}#2\end{tabular}}
	\renewcommand\arraystretch{1.5}
	\resizebox{\textwidth}{!}{
		\begin{tabular}{|c|c|c|c|c|}
			\hline
			\hline
			link & {intermediate variable}&clock noise&\makecell[c]{form in clock-noise observables}&type \\
			\hline
			$1$&$\tilde \eta^q _{13}$&$-\tilde a _{13}{q_1} + \tilde a _{13}{q_1}$&$0$&\\
			\hline
			$2$&${\cal D}_{13}\tilde \eta^q _{32}$&$-\tilde a _{32}{\cal D}_{13}{q_3} + \tilde a _{32}{q_1}$&$-\tilde a _{32}r_{13}$&$a$ \\
			\hline
			$3$&${\cal D}_{13}{\cal D} _{32}\tilde \eta^q _{21}$&$-\tilde a _{21}{\cal D}_{13}{\cal D} _{32}{q_2} + \tilde a _{21}{q_1}$&$-\tilde a _{21}({\cal D}_{13}r_{32}+r_{13})$&$a$ \\
			\hline
			$4$&${\cal D}_{13}{\cal D} _{32}{\cal D} _{21}\tilde \eta^q _{13}$&$-\tilde a _{13}{\cal D}_{13}{\cal D} _{32}{\cal D} _{21}{q_1} + \tilde a _{13}{q_1}$&$-\tilde a _{13}({\cal D}_{13}{\cal D} _{32}r_{21}+{\cal D}_{13}r_{32}+r_{13})$&$a$ \\
			\hline
			$5$&${\cal D}_{13}{\cal D} _{32}{\cal D} _{21}{\cal D} _{13}\tilde \eta^q _{32}$&$-\tilde a _{32}{\cal D}_{13}{\cal D} _{32}{\cal D} _{21}{\cal D} _{13}{q_3} +\tilde a _{32}{q_1}$&\makecell[c]{$-\tilde a _{32}\left( \begin{array}{l}{\cal D}_{13}{\cal D} _{32}{\cal D} _{21}r_{13}\\
					+{\cal D}_{13}{\cal D} _{32}r_{21}+{\cal D}_{13}r_{32}+r_{13}\end{array} \right)$}&$a$ \\
			\hline
			$6$&$-{\cal D}_{13}{\cal D} _{32}{\cal D} _{21}{\cal D} _{13}{\cal D}_{32}{\cal D}_{-21}\tilde \eta^q_{12}$&$\makecell[c]{\tilde a _{12}{\cal D}_{13}{\cal D} _{32}{\cal D} _{21}{\cal D} _{13}{\cal D}_{32}{\cal D}_{-21}{q_1} \\- \tilde a _{12}{q_1}}$&$\makecell[c]{\tilde a _{12}\left( \begin{array}{l}
					-{\cal D}_{13}{\cal D} _{32}{\cal D} _{21}{\cal D} _{13}{\cal D}_{32}{\cal D}_{-21}r_{12}\\
					+{\cal D}_{13}{\cal D} _{32}{\cal D} _{21}{\cal D}_{13}r_{32}\\
					{\cal D}_{13}{\cal D} _{32}{\cal D} _{21}r_{13}\\
					+{\cal D}_{13}{\cal D} _{32}r_{21}+{\cal D}_{13}r_{32}+r_{13}\end{array} \right)}$ &$b$\\
			\hline
			$7$&$-{\cal D}_{13}{\cal D} _{32}{\cal D} _{21}{\cal D} _{13}{\cal D}_{32}{\cal D}_{-21}{\cal D}_{-12}\tilde \eta^q_{21}$&$\makecell[c]{\tilde a _{21}{\cal D}_{13}{\cal D} _{32}{\cal D} _{21}{\cal D} _{13}{\cal D}_{32}{\cal D}_{-21}{\cal D}_{-12}{q_2} \\- \tilde a _{21}{q_1}}$&$\makecell[c]{\tilde a _{21}\left( \begin{array}{l}
					-{\cal D}_{13}{\cal D} _{32}{\cal D} _{21}{\cal D} _{13}{\cal D}_{32}{\cal D}_{-21}{\cal D}_{-12}r_{21}\\
					-{\cal D}_{13}{\cal D} _{32}{\cal D} _{21}{\cal D} _{13}{\cal D}_{32}{\cal D}_{-21}r_{12}\\
					+{\cal D}_{13}{\cal D} _{32}{\cal D} _{21}{\cal D}_{13}r_{32}\\
					{\cal D}_{13}{\cal D} _{32}{\cal D} _{21}r_{13}\\
					+{\cal D}_{13}{\cal D} _{32}r_{21}+{\cal D}_{13}r_{32}+r_{13}\end{array} \right)}$&$d$ \\
			\hline
			$8$&$\makecell[c]{-{\cal D}_{13}{\cal D} _{32}{\cal D} _{21}{\cal D} _{13}{\cal D}_{32}\\{\cal D}_{-21}{\cal D}_{-12}{\cal D}_{-23}\tilde \eta^q_{32}}$&$\makecell[c]{\tilde a _{32}{\cal D}_{13}{\cal D} _{32}{\cal D} _{21}{\cal D} _{13}{\cal D}_{32}\\{\cal D}_{-21}{\cal D}_{-12}{\cal D}_{-23}{q_3} \\- \tilde a _{32}{q_1}}$&$\makecell[c]{\tilde a _{32}\left( \begin{array}{l}
					-{\cal D}_{13}{\cal D} _{32}{\cal D} _{21}{\cal D} _{13}{\cal D}_{32}{\cal D}_{-21}{\cal D}_{-12}{\cal D}_{-23}r_{32}\\
					-{\cal D}_{13}{\cal D} _{32}{\cal D} _{21}{\cal D} _{13}{\cal D}_{32}{\cal D}_{-21}{\cal D}_{-12}r_{21}\\
					-{\cal D}_{13}{\cal D} _{32}{\cal D} _{21}{\cal D} _{13}{\cal D}_{32}{\cal D}_{-21}r_{12}\\
					+{\cal D}_{13}{\cal D} _{32}{\cal D} _{21}{\cal D}_{13}r_{32}\\
					{\cal D}_{13}{\cal D} _{32}{\cal D} _{21}r_{13}\\
					+{\cal D}_{13}{\cal D} _{32}r_{21}+{\cal D}_{13}r_{32}+r_{13}\end{array} \right)}$&$d$ \\
			\hline
			\hline
		\end{tabular}
	}
	\label{suppressclockblue}
\end{table}

Our third example is the geometric TDI combination $\left[ {\rm{U}} \right]_3^{16}$, whose space-time diagram is shown in the right panel of Fig.~\ref{fig5}.
The specific forms of relevant quantities for clock-noise reduction scheme are presented in Tabs.~\ref{suppressclockred}-\ref{suppressclockblue} for individual links.
The sum of the zeroth terms is found to be vanished,
\begin{align}\label{xiq10}
	&\left( \tilde a _{12} +\tilde a _{21} + \tilde a_{13} + \tilde a _{32}- \tilde a _{12} - \tilde a _{21} - \tilde a _{12} + \tilde a _{13} \right){q_1} \nonumber\\
	&- \left(   \tilde a _{13}+ \tilde a _{32}+ \tilde a _{21} + \tilde a _{13} + \tilde a _{32}- \tilde a _{12} - \tilde a _{21} -\tilde a _{32} \right){q_1}=0 ,
\end{align}
consistent with Eq.~\eqref{xicondition}.

\section{Numerical Simulations}\label{section5}

In this section, we present the performance of the suppression scheme using numerical simulations. 
As detailed below, we simulate gravitational wave signals along with various types of relevant noises using a given TDI solution.
Specifically, we apply the clock-noise reduction scheme to assess the effectiveness of our proposed formalism. 
For this numerical study, we choose the TDI combination $\left[U\right]_3^{16}$.

\begin{figure}[H]
	\centering
	\includegraphics[width=0.62\textwidth]{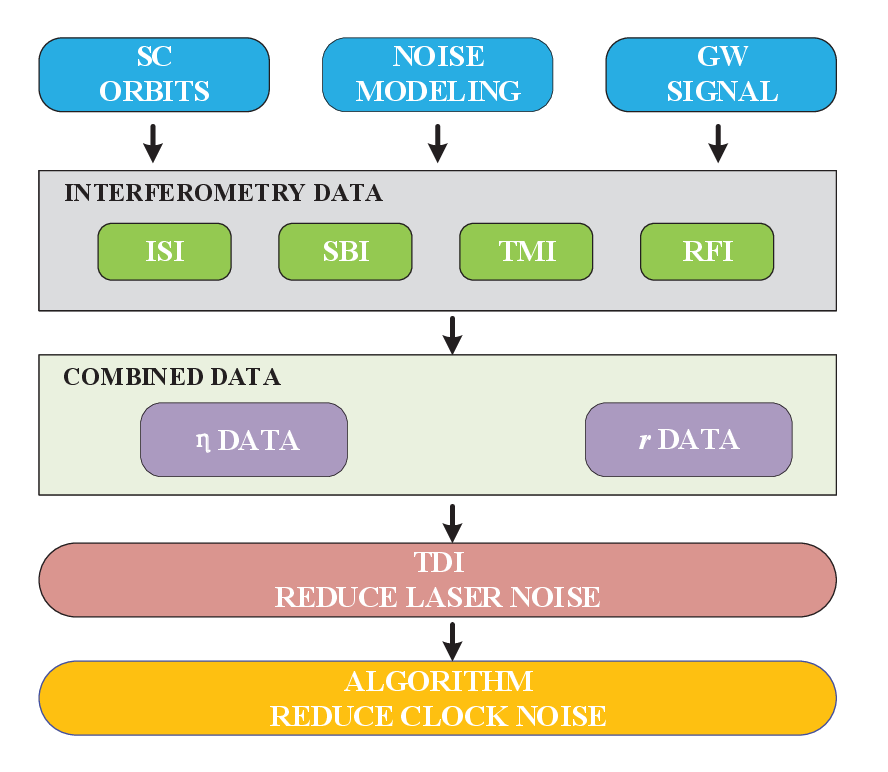}
	\caption{\label{fig6} 
		Flowchart of the numerical simulations for the laser phase and clock-noise reduction scheme. 
		ISI: interspacecraft interferometer (scientific carrier data stream); SBI: sideband interferometer (scientific sideband data stream); TMI: test-mass interferometer (test mass data stream); RFI: reference interferometer (reference data stream).}
\end{figure}

The flowchart of the numerical simulation is depicted in Fig.~\ref{fig6}, consisting of four main stages:
\begin{itemize}
	\item Generation of GW signals and various types of noise, for which the specific orbits of LISA's three spacecraft are considered when necessary.
	\item Derivation of relevant measurements based on the simulation results, obtaining data streams.
	\item Combination of the data streams using time-shift operations to calculate $\eta_{ij}$ and $r_{ij}$ as defined in Eqs.~\eqref{etaijik},~\eqref{rrij},~\eqref{r-ijdata},~\eqref{ri-jdata}, and~\eqref{r-i-jdata} in earlier sections.
	\item Implementation of the TDI algorithm in conjunction with the noise reduction scheme.
\end{itemize}

The relative frequency shift of the laser beam along a single arm due to the presence of GW in the time domain is given by~\cite{doppler-01}:
\begin{align}\label{N13}
	\frac{\dot h_{ij}(t)}{2\pi\nu_0} = \frac{- 1}{2(1 - \hat k \cdot \hat n)}\left[ {h(t - \hat k \cdot \frac{\vec r_j}{c}) - h(t - \hat k \cdot \frac{\vec r_i}{c} - \frac{L}{c})} \right],
\end{align}
where $\nu_0$ is the central frequency of the laser, $\hat k$ represents the direction of GW's wave vector. 
The two events $(t_i, \vec r_i)$ and $(t_j, \vec r_j)$ indicate those when a photon is emitted from spacecraft $i$ and received by $j$, $\hat n$ denotes the unit vector pointing in the direction of transmission $\vec r_i\to \vec r_j$, $L$ represents the Euclidean distance between the two spacecraft, and $c$ is the speed of light.
By performing the Fourier transform, one can rewrite the above result in the frequency domain as follows
\begin{align}\label{DopplerShift}
	\frac{h_{ij}(u )u}{2\pi\nu _0 \frac{L}{c}} = \frac{\tilde h(u )}{2(1 - \hat k \cdot \hat n)}{e^{iu\frac{L + \hat k \cdot {{\vec r}_i}}{c}}}\left[ {1 - {e^{ - iu(1 - \hat k \cdot \hat n)}}} \right],
\end{align}
where $u  = \frac{2\pi f L}{c}$ is dimensionless quantity.

Various noises are added on the top of GW signals.
A typical approach would be to generate some white noise and modulate it by a bandpass to achieve the desired frequency characteristics tailored to the specific noise source.
In our simulations, we explicitly consider the laser phase noise, clock noise, test mass acceleration noise, and optical path noise.
The strengths and frequency dependence of these noises are determined by typical LISA parameters.
In particular, the laser frequency noise is given by~\cite{unitmodelsim-2023}:
\begin{align}\label{lasernoise}
	{s_p} = 30\left( {\sqrt {1 + {{\left( {\frac{{2{\rm{mHz}}}}{f}} \right)}^4}} } \right){\rm{Hz/}}\sqrt {{\rm{Hz}}},
\end{align}
The clock noise assumes the form~\cite{tdi-clock-2021}:
\begin{align}\label{clocknoise}
	{s_q} = \frac{\delta f}{f_0} \sim 6.32 \times {10}^{ - 14}\sqrt{\frac{1\rm{Hz}}{ f }}\sqrt {\rm{Hz}}.
\end{align}
Besides, the optical path noise and test mass acceleration noise are governed, respectively, by the forms~\cite{sasxdef-2021}
\begin{align}\label{optnoise}
	{s_x} = 15\left( {\sqrt {1 + {{\left( {\frac{2 \times 10^{ - 3}\rm{Hz}}{f}} \right)}^4}} } \right){\rm{pm/}}\sqrt {{\rm{Hz}}},
\end{align}
and
\begin{align}\label{testnoise}
	&{s_a} = 3 \times {10^{ - 15}}\nonumber\\
	&\times\sqrt {1 + {{\left( {\frac{{0.4 \times {\rm{1}}{{\rm{0}}^{ - 3}}}\rm{Hz}}{f}} \right)}^2}} \sqrt {1 + {{\left( {\frac{f}{{8 \times {{10}^{ - 3}\rm{Hz}}}}} \right)}^4}} {\rm{m}} \cdot {{\rm{s}}^2}{\rm{/}}\sqrt {{\rm{Hz}}}.
\end{align}

For a realistic space-based detector scenario, measurements involve three arms through which laser beams are transmitted in both directions. 
We utilize the information on spacecraft orbits from~\cite{ORBIT-LISA-2021}. 
The values of $\vec r_i(t)$, $\hat{n}_i$, and $L_i$ where $i=1,2,3$ are updated accordingly, while their time dependence is ignored when deriving the Fourier-transformed result given in Eq.~\eqref{DopplerShift}.

We decompose the GW into plus and cross modes, ${\tilde h_ + }(\Omega )$ and ${\tilde h_ \times }(\Omega )$, and the response functions $\tilde F_ +$ and $\tilde F_ \times$ are determined by the specific detector layout concerning the two polarization states~\cite{gravity-2001}. 
The strains of a monochromatic GW source are given by~\cite{gravity-1980}:
\begin{align}\label{hexpression}
	\left\{ \begin{array}{l}
		{h_{\rm{ + }}}(t) \!\equiv\! H\left[ {\frac{{1 +\cos^2\iota }}{2}\cos 2\psi \cos 2\pi f_{GW} t \!+\! \cos \iota \sin 2\psi \sin 2\pi f_{GW} t} \right]{\rm{  }},\\
		{h_{\rm{ \times }}}(t) \!\equiv\! H\left[ { \!- \frac{{1 \!+\! \cos^2\iota }}{2}\sin 2\psi \cos 2\pi f_{GW} t \!+\! \cos \iota \cos 2\psi \sin 2\pi f_{GW} t} \right],
	\end{array} \right.
\end{align}
and in particular, for a binary system, $H = \frac{{4{G^2}{m_1}{m_2}}}{{{c^4}Ra}}$ represents the amplitude of GWs, where $G$ is the gravitational constant, $m_1$ and $m_2$ are the masses of the binary, $R$ is the distance from the source, and $a$ measures spatial separation between the two compact objects.
The inclination $\iota$ measures the angle between the plane of the source's orbit and that perpendicular to the GW's propagation.
The polarization angle is $\psi$.
In our simulation, for simplicity, the inclination angle is taken as $\iota=0$.
We have confirmed through further analysis that the primary results are consistent across different inclination angles.
For a given TDI combination, the response function to the GW signal for the plus and cross modes are
\begin{align}\label{plus}
	R{\left( u \right)_ + } =  - \frac{1}{2}\left( {1 - {e^{iu}}} \right)\left[ {\frac{1}{2}({\tilde P_{12}}  + {\tilde P_{31}} + {\tilde P_{13}} + {\tilde P_{21}}) - ({\tilde P_{23}} + {\tilde P_{32}}}) \right],
\end{align}
and
\begin{align}\label{cross}
	R{\left( u \right)_ \times } = \frac{{\sqrt 3 }}{4}\left( {1 - {e^{iu}}} \right)({\tilde P_{12}} - {\tilde P_{31}} - {\tilde P_{13}} + {\tilde P_{21}}),
\end{align}
where $e^{iu}$ is the Fourier transform of the time-delay operator, $\tilde P_{ij}$, $\tilde P_{ik}$ are the Fourier transform of the polynomial coefficients in time delay operators associated with the TDI combination.
The resulting response function reads
\begin{align}\label{Ru}
	R(u) = R{\left( u \right)_ + }{\tilde h_ + }\left( u \right) + R{\left( u \right)_ \times }{\tilde h_ \times }\left( u \right).
\end{align}
Finally the sensitivity of the detector regarding the input GW signal reads
\begin{align}\label{Sensitivity}
	S(u)=\sqrt{\frac{N(u)}{R(u)}} .
\end{align}
where $N(u)$ is the detector noise floor.

By taking into account the TDI algorithm, the resultant clock noise's power spectral density (PSD) is given by
\begin{align}\label{clocknoisePSD}
	&{N_{{\rm{TD}}{{\rm{I}}^q}}}(u ) \approx\nonumber\\
	&{S_q}(u ){\sum\limits_{i = 1}^3 {\left| {{a_{ij}}{{\tilde P}_{ij}}(u ) + {a_{ik}}{{\tilde P}_{ik}}(u ) - {b_{ik}}\left[ {{{\tilde P}_{ik}}(u ) - {{\tilde P}_{ki}}(u){{\tilde D}_{jk}}(u)} \right]} \right|} ^2}.
\end{align}
where flicker noise is $S_q={s_q}^2$.
The resulting PSD of test mass noise is
\begin{align}\label{testnoisePSD}
	&{N_{{\rm{TD}}{{\rm{I}}^a}}}(u) =\nonumber\\
	&\frac{{s_a^2{L^2}}}{{{u^2}{c^2}}}\sum\limits_{i,j,k \in \Gamma _3^ + }^3 {\left[ {{{\left| {{{\tilde P}_{ij}}(u) + {{\tilde P}_{ji}}(u ){{\tilde{\cal D}_{kj}}}\left( u \right)} \right|}^{\rm{2}}} + {{\left| {{{\tilde P}_{ij}}(u){{\tilde{\cal D}_{ki}}}\left( u \right) + {{\tilde P}_{ji}}(u)} \right|}^{\rm{2}}}} \right]},
\end{align}
and the PSD of the optical path noise reads
\begin{align}\label{shotnoisePSD}
	{N_{{\rm{TD}}{{\rm{I}}^x}}}{\rm{(}}u{\rm{) = }}\frac{{{u^2}s_x^2}}{{{L^2}{c^2}}}\sum\limits_{i = 1}^3 {\left[ {{{\left| {{{\tilde P}_{ij}}\left( u  \right)} \right|}^2} + {{\left| {{{\tilde P}_{ik}}(u )} \right|}^2}} \right]},
\end{align}
where $s_a$ is the amplitude spectral density (ASD) of test mass acceleration noise, and $s_x$ is the ASD of displacement noise.

By implementing the simulation including specific information on the spacecraft's orbits, GW signal, and relevant noises, the data streams, namely, the scientific carrier, scientific sideband, test mass, and reference data streams defined in Eqs.~\eqref{sci}-\eqref{ref}, are generated as time series.
Following the standard procedure of the TDI algorithm and the clock-noise reduction scheme elaborated in the present study, the data streams are combined to derive the variables $\eta_{ij}$ and $r_{ij}$ defined in Eqs.~\eqref{etaijik},~\eqref{rrij},~\eqref{r-ijdata},~\eqref{ri-jdata}, and~\eqref{r-i-jdata}.

For illustration purpose, we only consider the TDI combination $\left[ U \right]_{\rm{3}}^{{\rm{16}}}$ combination, whose polynomial coefficients in time delay operators are~\cite{geome-tdi-2023}
\begin{align}\label{polyu316}
	{P_{{\rm{12}}}} =& 1 - {{\cal D}_{12}}{{\cal D}_{2{\rm{1}}}}{{\cal D}_{{\rm{13}}}}{{\cal D}_{{\rm{32}}}}{{\cal D}_{ - 21}} - {{\cal D}_{12}}{{\cal D}_{2{\rm{1}}}}{{\cal D}_{{\rm{13}}}}{{\cal D}_{{\rm{32}}}}{{\cal D}_{ - 21}}{{\cal D}_{ - 12}}{{\cal D}_{ - 21}}\notag\\
	+& {{\cal D}_{13}}{{\cal D}_{32}}{{\cal D}_{21}}{{\cal D}_{13}}{{\cal D}_{32}}{{\cal D}_{ - 21}} ,\\\notag
	{P_{{\rm{23}}}} =& {\rm{0}},\\\notag
	{P_{{\rm{31}}}} =& {\rm{0}},\\\notag
	{P_{13}} =& - 1 + {{\cal D}_{12}}{{\cal D}_{2{\rm{1}}}} - {{\cal D}_{13}}{{\cal D}_{32}}{{\cal D}_{21}}\notag\\
	+& {{\cal D}_{12}}{{\cal D}_{2{\rm{1}}}}{{\cal D}_{{\rm{13}}}}{{\cal D}_{{\rm{32}}}}{{\cal D}_{ - 21}}{{\cal D}_{ - 12}}{{\cal D}_{ - 21}}{{\cal D}_{ - 21}} ,\\\notag
	{P_{{\rm{21}}}} =&  {{\cal D}_{12}} - {{\cal D}_{13}}{{\cal D}_{32}} - {{\cal D}_{12}}{{\cal D}_{2{\rm{1}}}}{{\cal D}_{{\rm{13}}}}{{\cal D}_{{\rm{32}}}}{{\cal D}_{ - 21}}{{\cal D}_{ - 12}}\notag\\
	+& {{\cal D}_{13}}{{\cal D}_{32}}{{\cal D}_{21}}{{\cal D}_{13}}{{\cal D}_{32}}{{\cal D}_{ - 21}}{{\cal D}_{ - 12}} ,\\\notag
	{P_{{\rm{32}}}} =&   - {{\cal D}_{13}} + {{\cal D}_{12}}{{\cal D}_{2{\rm{1}}}}{{\cal D}_{{\rm{13}}}} - {{\cal D}_{13}}{{\cal D}_{32}}{{\cal D}_{21}}{{\cal D}_{13}}\notag\\
	+& {{\cal D}_{13}}{{\cal D}_{32}}{{\cal D}_{21}}{{\cal D}_{13}}{{\cal D}_{32}}{{\cal D}_{ - 21}}{{\cal D}_{ - 12}}{{\cal D}_{ - 23}}.
\end{align}
As the present study primarily concerns clock-noise reduction, we focus on the numerical results demonstrating the effectiveness of the proposed scheme.
For both the simulations and theoretical values, we note that the response function is given by Eq.~\eqref{Ru}, and the relevant noises are given by Eqs.~\eqref{lasernoise},~\eqref{clocknoise},~\eqref{optnoise}, and~\eqref{testnoise}.
They are obtained by substituting the simulated time series and the underlying analytic expressions.
Nonetheless, theoretical values for the noise PSDs can be obtained analytically and given by Eqs.~\eqref{clocknoisePSD},~\eqref{testnoisePSD}, and~\eqref{shotnoisePSD}.

For the numerical simulations, the central frequency of the laser is taken as ${\nu _0} \sim 3 \times {10^{14}}{\rm{Hz}}$, and the local frequency of the clock is ${f_0} \sim 10 \times {10^6}{\rm{Hz}}$.
Also, the beat frequency coefficients are set to $5-20{\rm{MHz}}$, the detector arm length $L=2.5\times 10^9$.
Besides the above typical LISA parameters, the simulated time series are interpolated using 20th-order Lagrange fractional delay filters before further processing.
We take simulated data for a span of observation that lasted for two days.
Regarding the scientific goal of the space-based GW projects, the magnitudes of the laser phase noise, optical bench noise, and clock noise are taken to be more significant than that of the GW signals.
In this regard, the amplitude of the GW signal in Eq.~\eqref{hexpression} is taken to be $H = \frac{1 \times 10^{ - 18}}{\sqrt{\rm Hz}}\sqrt {\frac{2}{T}}$ with a frequency $f_\mathrm{GW}=1{\rm{mHz}}$, $T=2 \rm{days}$.

\begin{figure}[H]
	\centering
	\includegraphics[width=0.62\textwidth]{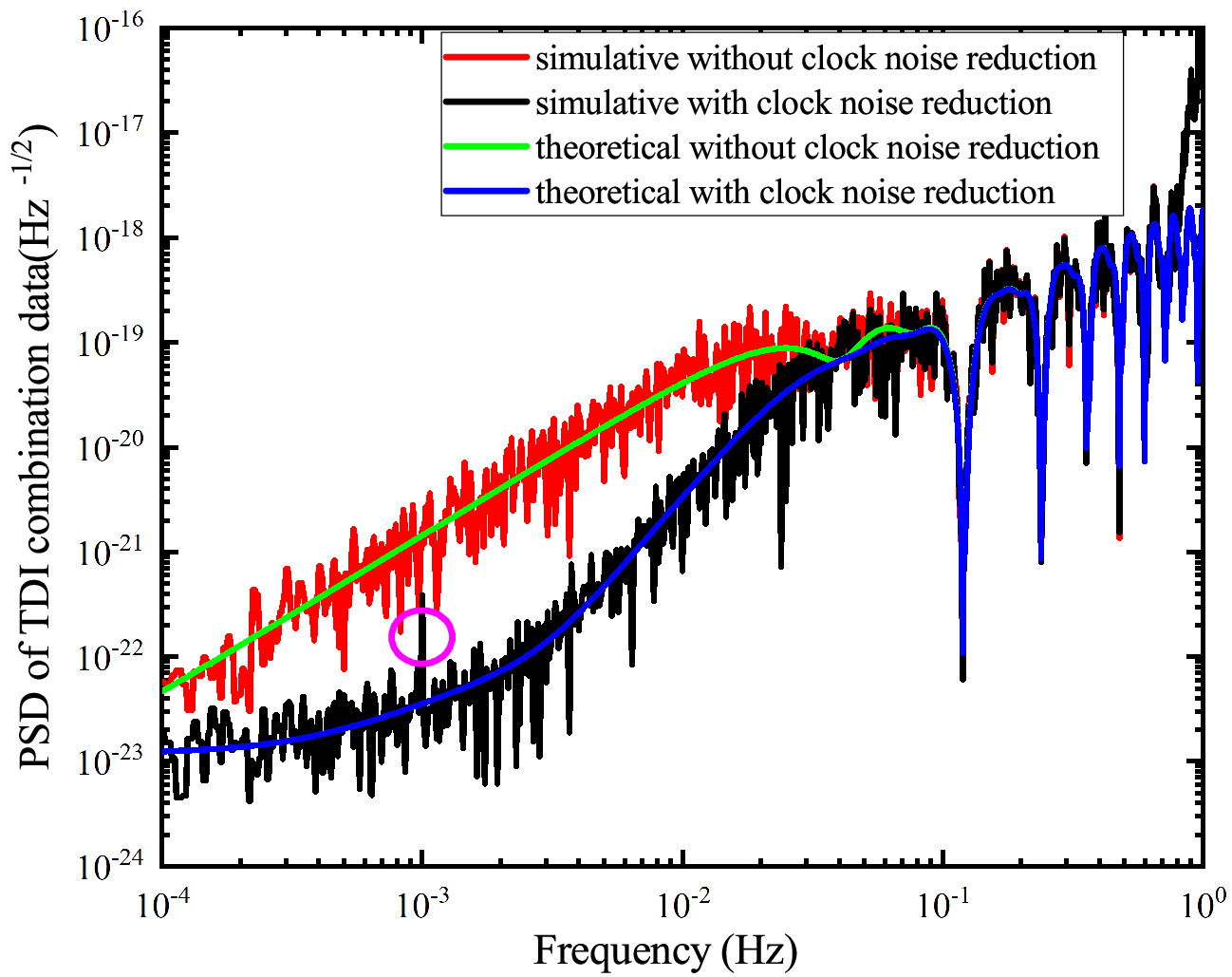}
	\caption{\label{fig7}
		The residual noise PSD of a monochromatic GW source located at the north celestial pole (w.r.t. the detector plane) when the TDI combination $[U]_3^{16}$ is applied.
		The curves correspond to scenarios before and after implementing the proposed clock-noise suppression algorithm. 
		The red and green curves represent the noise residual after applying the TDI algorithm but not the clock-noise reduction procedure.
		The black and blue curves are the resulting PSD when the clock noise is further suppressed. 
		The red and black curves show the simulation results, while the green and blue curves are the corresponding theoretical values.
		The highlighted peak enclosed by the empty magenta circle indicates the GW signal.}
\end{figure}

\begin{figure}[H]
	\centering
	\includegraphics[width=0.62\textwidth]{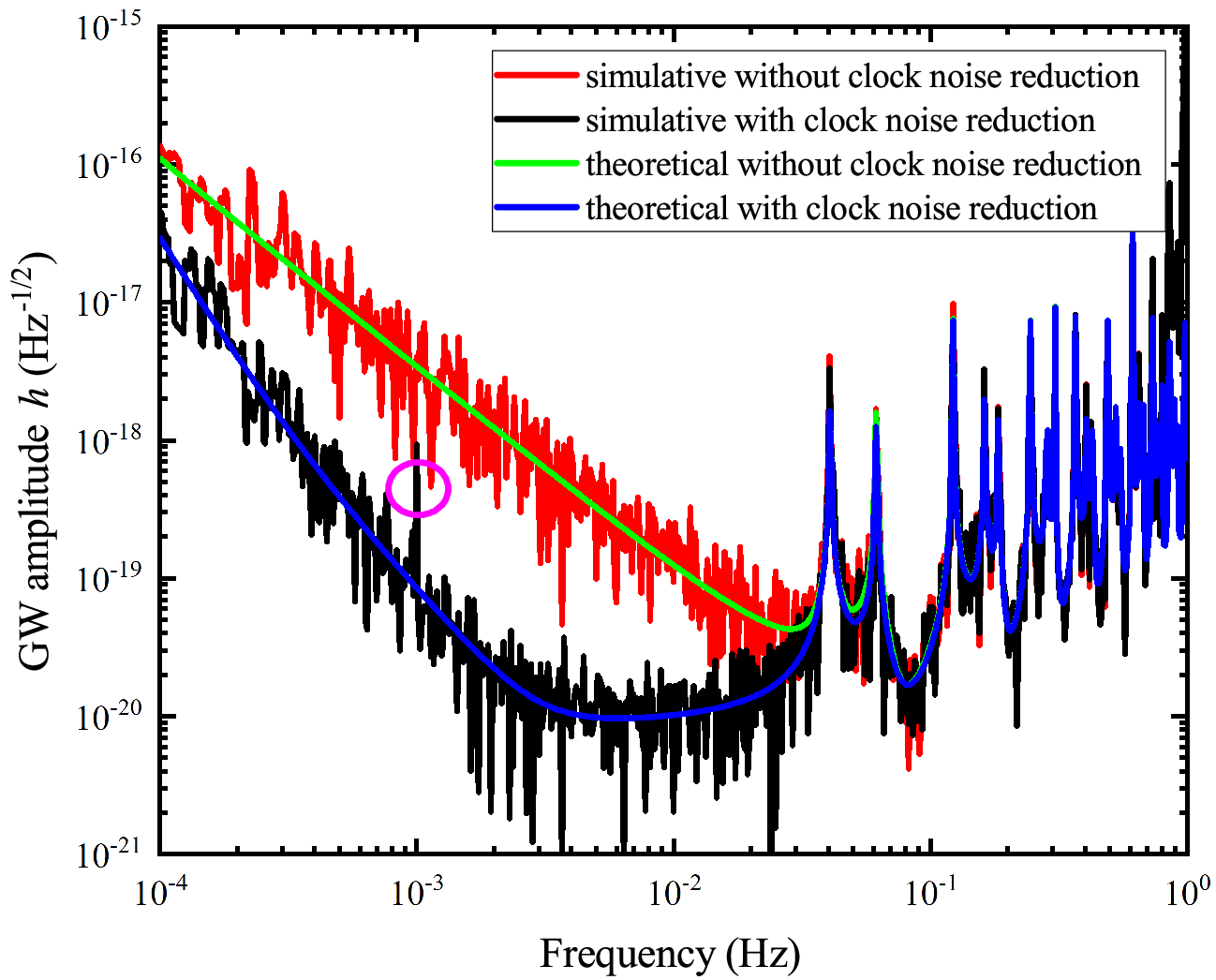}
	\caption{\label{fig8}
		The sensitivity curve of a monochromatic GW source located at the north celestial pole (w.r.t. the detector plane) when the TDI combination $[U]_3^{16}$ is applied.
		The curves correspond to scenarios before and after implementing the proposed clock-noise suppression algorithm. 
		The legend of the plot is the same as Fig.~\ref{fig7}.}
\end{figure}

The numerical results are presented in Figs.~\ref{fig7} and~\ref{fig8} for the residual noise's PSD and the sensitivity curve.
The red and green curves represent the simulated and theoretical results after applying the TDI algorithm but not any clock-noise reduction.
The black and blue curves depict the simulated and theoretical results after implementing the proposed clock-noise suppression scheme.
One observes that satisfactory agreement between the theoretical estimation and numerical simulation is achieved.
Moreover, even though laser phase noise can be readily suppressed below the noise floor level, it is evident that the residual noise indicated by the green and red curves overwhelms the GW signal. 
By employing the algorithm elaborated in Sec.~\ref{section4.1}, the clock-noise residual is also manifestly suppressed below the signal, demonstrating the algorithm's effectiveness.

\section{Parameter estimation for TDI combinations}\label{section6}

The ultimate purpose of instrumental-noise suppression is not only to improve sensitivity curves, but also to preserve the information required to infer the physical parameters of GW sources.
For space-based detectors, parameter estimation is performed on TDI observables rather than on individual inter-spacecraft measurements.
Any residual instrumental contribution that remains in a TDI channel changes the effective noise weighting in the likelihood and therefore affects the posterior width of the recovered source parameters.
Clock-noise subtraction is consequently relevant for data analysis: once the laser phase noise has been removed by TDI, uncompensated clock noise can still degrade the precision with which amplitudes, frequencies and phases are measured.

In this section we quantify this effect with the modified second-generation geometric TDI observable $[U]_3^{16}$ studied in Sec.~\ref{section5}.
The numerical simulations above show that the proposed subtraction scheme reduces the residual clock noise of this observable below the secondary-noise floor around the injected monochromatic signal.
We now use the same LISA-like noise model and the same $[U]_3^{16}$ response to compare the parameter-estimation performance before and after clock-noise subtraction.

We consider a monochromatic source with the waveform model introduced in Eq.~\eqref{hexpression}.
For the baseline analysis the sky location, inclination and polarization angle are fixed to the values used in the numerical simulation, and we estimate the three intrinsic parameters
\begin{align}
	\bm{\theta} = \left(\ln H,\ f_{\rm GW},\ \phi_0\right),
	\label{eq:pe_parameters_u316}
\end{align}
where $H$ is the strain amplitude, $f_{\rm GW}$ is the GW frequency and $\phi_0$ is the initial phase.
The data stream is modeled as
\begin{align}
	d_U(t) = h_U(t;\bm{\theta}_{\rm inj}) + n_U(t),
	\label{eq:pe_data_model_u316}
\end{align}
where $h_U(t;\bm{\theta})$ denotes the response of the $[U]_3^{16}$ TDI combination and $n_U(t)$ is stationary Gaussian noise.
The likelihood is written in the standard form
\begin{align}
	\ln {\cal L}(\bm{\theta})
	=
	-\frac{1}{2}
	\left[
	d_U-h_U(\bm{\theta})
	\middle|
	d_U-h_U(\bm{\theta})
	\right],
	\label{eq:pe_likelihood_u316}
\end{align}
with the noise-weighted inner product
\begin{align}
	(a|b)
	=
	4\,{\rm Re}\int_0^\infty
	\frac{\tilde a^\ast(f)\tilde b(f)}{S_U(f)}\,df .
	\label{eq:pe_inner_product_u316}
\end{align}
In the present single-source demonstration this expression is evaluated in the time domain using the noise level at the injected frequency,
\begin{align}
	\ln {\cal L}(\bm{\theta})
	\simeq
	-\frac{\Delta t}{S_U(f_{\rm GW})}
	\sum_n
	\left[
	d_U(t_n)-h_U(t_n;\bm{\theta})
	\right]^2 .
	\label{eq:pe_time_likelihood_u316}
\end{align}
The two cases compared below differ only in the noise PSD used in the likelihood.
Before clock-noise subtraction we take
\begin{align}
	S_U^{\rm before}(f)
	=
	S_U^{\rm sec}(f)+S_U^{\rm clk}(f),
	\label{eq:pe_noise_before_u316}
\end{align}
where $S_U^{\rm sec}$ contains the test-mass and optical-path contributions and $S_U^{\rm clk}$ is the residual clock-noise PSD.
After applying the subtraction scheme, the clock contribution is replaced by the residual left by the calibration procedure,
\begin{align}
	S_U^{\rm after}(f)
	=
	S_U^{\rm sec}(f)+\epsilon_{\rm clk}S_U^{\rm clk}(f),
	\label{eq:pe_noise_after_u316}
\end{align}
where $\epsilon_{\rm clk}=0$ corresponds to ideal subtraction in the plots shown here.
The expressions for $S_U^{\rm clk}$, $S_U^{\rm sec}$ and the $[U]_3^{16}$ delay-polynomial coefficients are the same as those used in Sec.~\ref{section5}.

For the Bayesian analysis we adopt broad uniform priors around the injected values,
\begin{align}
	\ln H &\in [\ln H_{\rm inj}-5,\ \ln H_{\rm inj}+5],\nonumber\\
	f_{\rm GW} &\in
	\left[
	f_{\rm inj}-\frac{5}{T},\
	f_{\rm inj}+\frac{5}{T}
	\right],\nonumber\\
	\phi_0 &\in [0,2\pi],
	\label{eq:pe_priors_u316}
\end{align}
where $T=2~{\rm days}$ is the observation time.
The posterior is then
\begin{align}
	p(\bm{\theta}|d_U)
	\propto
	{\cal L}(\bm{\theta})\,
	\pi(\bm{\theta}) .
	\label{eq:pe_posterior_u316}
\end{align}
We sample this posterior with a Metropolis--Hastings Markov-chain Monte Carlo algorithm.
For display purposes the injected amplitude is normalized such that the post-subtraction signal-to-noise ratio is ${\rm SNR}=20$; using the amplitude adopted in Sec.~\ref{section5} changes the absolute widths of the posteriors but not the relative degradation caused by residual clock noise.

\begin{figure}[H]
	\centering
	\includegraphics[width=0.78\textwidth]{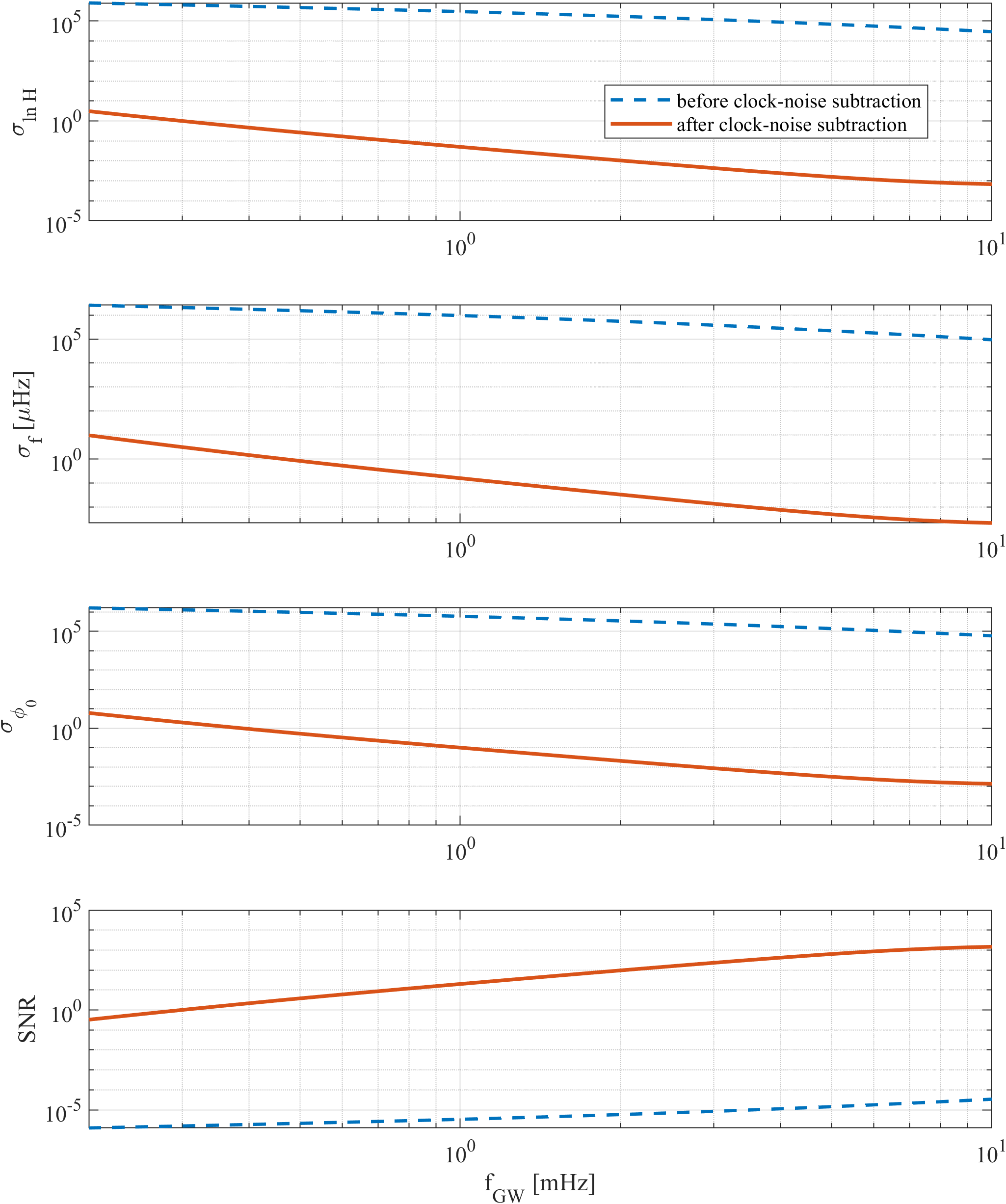}
	\caption{\label{fig:parameter_errors_u316}
	Parameter-estimation uncertainty for a monochromatic source observed with the $[U]_3^{16}$ TDI combination.
	The dashed curves are obtained before clock-noise subtraction, while the solid curves correspond to the clock-noise-subtracted case.
	The panels show the one-sigma uncertainties of $\ln H$, $f_{\rm GW}$ and $\phi_0$, together with the corresponding SNR, as functions of the injected GW frequency.
	Clock-noise subtraction lowers the effective noise PSD entering the likelihood and therefore improves the inferred source parameters.}
\end{figure}

\begin{figure}[H]
	\centering
	\includegraphics[width=0.78\textwidth]{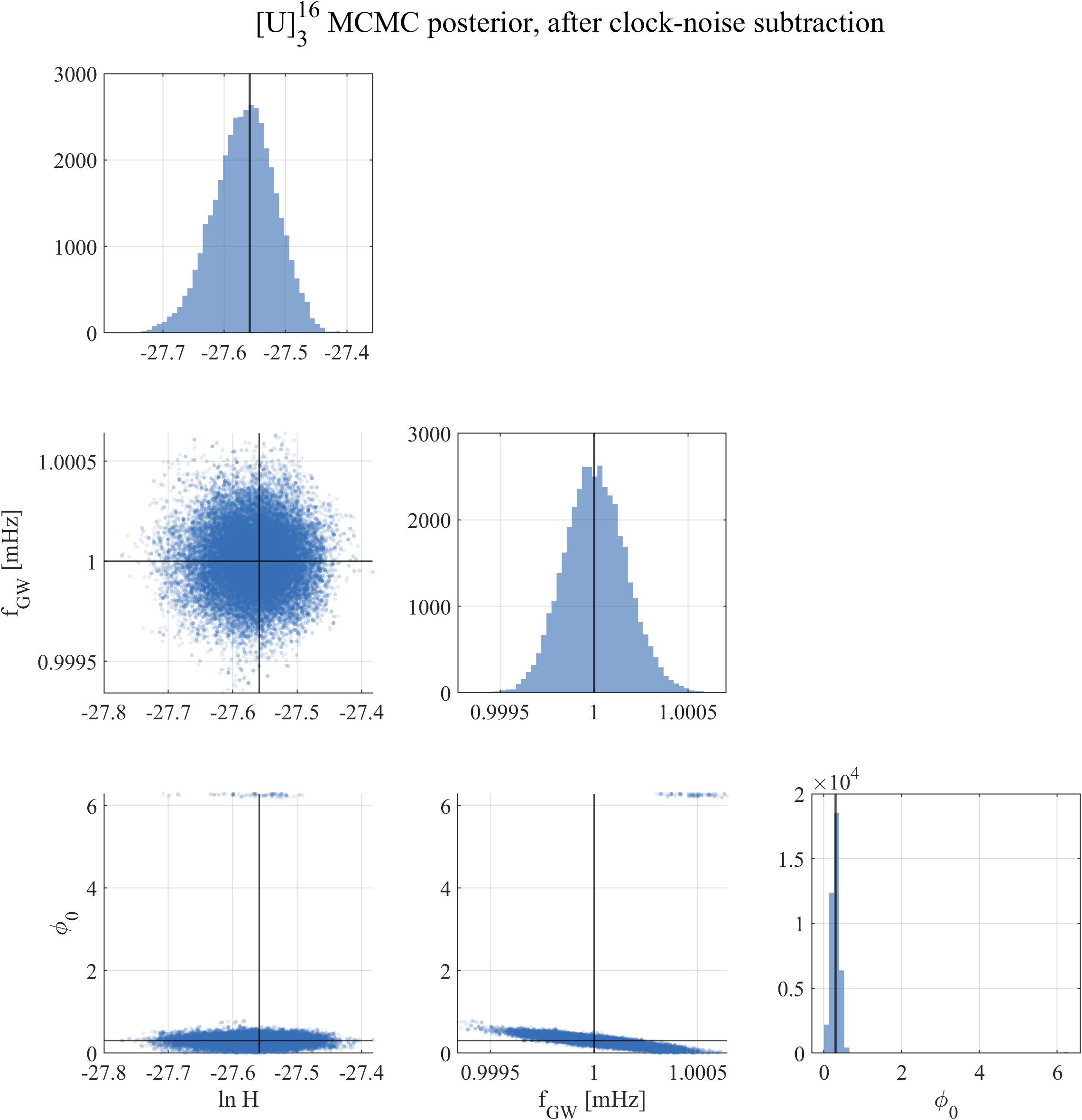}
	\caption{\label{fig:mcmc_corner_u316}
	MCMC posterior samples for the parameters $(\ln H,f_{\rm GW},\phi_0)$ after clock-noise subtraction.
	The vertical and horizontal black lines indicate the injected values.
	The posterior is localized around the injection, showing that the clock-noise-subtracted $[U]_3^{16}$ observable retains the phase and amplitude information required for parameter inference.}
\end{figure}

The comparison demonstrates that residual clock noise has a direct impact on parameter estimation.
When the clock-noise contribution is retained in $S_U(f)$, the effective noise entering Eq.~\eqref{eq:pe_likelihood_u316} is larger and the posterior volume increases.
After subtraction, the posterior contracts around the injected parameters, and the frequency and phase uncertainties are reduced together with the amplitude uncertainty.
This behavior is consistent with the sensitivity improvement found in Sec.~\ref{section5}: the same reduction of the noise floor that restores the visibility of the monochromatic signal also improves the precision with which the signal parameters can be recovered.

\section{Conclusions}\label{section7}

We have developed a clock-noise subtraction scheme for geometric TDI and assessed its impact on both sensitivity recovery and parameter estimation for space-based GW detectors.
The motivation is that TDI removes the dominant laser phase noise, but it does not by itself guarantee that residual clock jitter from onboard ultra-stable oscillators is below the secondary-noise floor.
For LISA-like parameters this residual can affect not only the visibility of millihertz GW signals, but also the likelihood weighting used in source-parameter inference.

The main result of this work is a geometric formulation of clock-noise subtraction that applies to arbitrary two-path geometric TDI observables, including combinations with time-advance operators.
We introduced generalized clock-noise observables associated with the four possible space-time link structures.
These observables allow the clock-noise residual to be written in a form directly analogous to the laser-noise residual in geometric TDI.
As a consequence, the subtraction terms can be constructed from measured carrier, sideband, test-mass and reference interferometric data streams without changing the underlying geometric-TDI logic.

We illustrated the construction with the Monitor-E, modified second-generation Michelson and modified second-generation U-type observables.
For each case the required time-shifted clock-noise observables were given explicitly.
We then tested the method in time-domain simulations with LISA-like orbits and noise levels.
For the U-type observable considered in detail, the proposed subtraction suppresses the residual clock-noise contribution and restores the expected sensitivity to a monochromatic GW signal.
A follow-up parameter-estimation study shows the corresponding data-analysis consequence: after clock-noise subtraction, the effective noise PSD entering the likelihood is reduced, the signal-to-noise ratio increases and the posterior constraints on the source amplitude, frequency and phase become more localized.

These results indicate that clock-noise calibration should be treated as part of the precision data-analysis chain for future space-based GW missions.

\section*{Data availability statement}
The numerical data and scripts supporting the parameter-estimation figures and the clock-noise subtraction tests are available from the corresponding authors upon reasonable request.

\acknowledgments
This work is supported by the National Key R$\&$D Program of China under Grants No.2022YFC2204602, the National Natural Science Foundation of China under Grants  No.12175076.

\paragraph{Declaration of AI-assisted technology.}
During the preparation of this manuscript, the authors used AI-assisted tools for language editing. The authors reviewed and edited all AI-assisted outputs and take full responsibility for the content of the manuscript.

\bibliographystyle{iopart-num}
\bibliography{references_luo}

\end{document}